\documentclass[]{aa}
\usepackage{epsfig}
\usepackage{natbib}
\bibpunct{(}{)}{;}{a}{}{,}
\def\etal{{\rm et al.\thinspace}}
\def\eg{{\rm e.g.\ }}

\def\ie{{\rm i.e.\ }}
\def\cf{{\rm cf.\ }}

\def\spose#1{\hbox to 0pt{#1\hss}}
\def\ltsimm{\mathrel{\spose{\lower 3pt\hbox{$\sim$}}
        \raise 2.0pt\hbox{$<$}}}
\def\gtsimm{\mathrel{\spose{\lower 3pt\hbox{$\sim$}}
        \raise 2.0pt\hbox{$>$}}}
\def\Mdot{\hbox{${\dot M}$}}

\def\km{{\rm\thinspace km}}
\def\cm{{\rm\thinspace cm}}
\def\s{{\rm\thinspace s}}
\def\yr{{\rm\thinspace yr}}
\def\kmps{\hbox{${\rm\km\s^{-1}\,}$}}

\def\erg{{\rm\thinspace erg}}
\def\Hz{{\rm\thinspace Hz}}
\def\ergps{\hbox{${\rm\erg\s^{-1}\,}$}}
\def\Msol{\hbox{${\rm\thinspace M_{\odot}}$}}
\def\Lsol{\hbox{${\rm\thinspace L_{\odot}}$}}
\def\Msolpyr{\hbox{${\rm\Msol\yr^{-1}\,}$}}

\def\ergpscm3Hz{\hbox{${\rm\ergps\cm^{-3}\Hz^{-1}\,}$}}

\value{1}

\begin{document}

\title{Radio emission models of Colliding-Wind Binary Systems}

\author{S.M.~Dougherty\inst{1}, J.M.~Pittard\inst{2}, L.~Kasian\inst{1}, 
       R.F.~Coker\inst{3}, P.M. Williams\inst{4} and H.M.~Lloyd\inst{5}}

\institute{National Research Council of Canada, Herzberg Institute for
        Astrophysics, Dominion Radio Astrophysical Observatory, P.O. Box
        248, Penticton, British Columbia V2A 6J9, Canada
        \and Department of Physics and Astronomy, The University of
        Leeds, Woodhouse Lane, Leeds LS2 9JT, UK \and Los Alamos
        National Laboratory, X-2 MS T-087, Los Alamos, NM 87545, USA
        \and Institute for Astronomy, University of Edinburgh, Royal
        Observatory, Blackford Hill, Edinburgh EH9 3HJ, UK \and Blade
        Interactive Studios, 274 Deansgate, Manchester M3 4JB, UK}

\offprints{S.M.~Dougherty,\\\email{sean.dougherty@nrc.gc.ca}}

\date{Origainally submitted April 24, 2003 / Re-submitted July 4, 2003 }

\abstract{We present calculations of the spatial and spectral
distribution of the radio emission from a wide WR+OB colliding-wind
binary system based on high-resolution hydrodynamical simulations and
solutions to the radiative transfer equation. We account for both
thermal and synchrotron radio emission, free-free absorption in both
the unshocked stellar wind envelopes and the shocked gas, synchrotron
self-absorption, and the Razin effect. To calculate the synchrotron
emission several simplifying assumptions are adopted: the relativistic
particle energy density is a simple fraction of the thermal particle
energy density, in equipartition with the magnetic energy density, and
a power-law in energy. We also assume that the magnetic field is
tangled such that the resulting emission is isotropic. The
applicability of these calculations to modelling radio images and
spectra of colliding-wind systems is demonstrated with models of the
radio emission from the wide WR+OB binary \object{WR\,147}. Its
synchrotron spectrum follows a power-law between 5 and 15~GHz but
turns down to below this at lower and higher frequencies. We find that
while free-free opacity from the circum-binary stellar winds can
potentially account for the low-frequency turnover, models that also
include a combination of synchrotron self-absorption and Razin effect
are favoured.  We argue that the high-frequency turn down is a
consequence of inverse-Compton cooling. We present our resulting
spectra and intensity distributions, along with simulated MERLIN
observations of these intensity distributions. From these we argue
that the inclination of the \object{WR\,147} system to the plane of
the sky is low. We summarise by considering extensions of the current
model that are important for models of the emission from closer
colliding wind binaries, in particular the dramatically varying radio
emission of \object{WR\,140}.   }

\titlerunning{Radio emission models of Colliding-Wind Binary Systems}
\authorrunning{S.M.~Dougherty et al.}

\maketitle

\label{firstpage}

\section{Introduction}
\label{sec:intro}
Observations of early-type stars have revealed that they can be sources of
both thermal and synchrotron radio emission. Thermal emission
typically exhibits a brightness temperature $\sim 10^{4}$~K and a spectral
index $\alpha\sim+0.6$ ($S_{\nu} \propto \nu^{\alpha}$) at centimetre
wavelengths, as expected from steady-state, radially
symmetric winds \citep{Wright:1975}. In contrast, synchrotron emission
is characterized by high brightness temperatures ($\ge 10^{6}$~K) and
flat or negative radio spectral indices. In addition to magnetic
fields, synchrotron emission requires a population of relativistic
electrons, which are widely thought to be accelerated in shocks. 
For single stars, shocks arise due to wind
instabilities \eg\citet{Lucy:1980}, and propagate through the wind,
while for massive binary systems stationary shocks arise where the
winds of the two stars collide \eg\citet{Eichler:1993}. Magnetic field
compression within the shocked gas is another potential acceleration
mechanism \citep{Jardine:1996}.

Spatially resolved observations of the WR+OB binary systems
\object{WR\,146} \citep{Dougherty:1996,Dougherty:2000a} and
\object{WR\,147} \citep{Moran:1989, Churchwell:1992, Williams:1997,
Niemela:1998} have dramatically confirmed the colliding-wind binary
(CWB) model in these objects.  In both \object{WR\,146} and
\object{WR\,147} the thermal emission is coincident with the position
of the WR star, while the synchrotron emission arises between the
binary components at a position consistent with the pressure balance
of the two stellar winds. This model is supported further by the
dramatic variations of the synchrotron radio emission in the 7.9-year
WR+OB system \object{WR\,140}, which are clearly modulated by the binary orbit
\eg~\citet{Williams:1990a,White:1995}.

Such results prompted ~\citet{vanderhucht:1992} to suggest a CWB
origin for all synchrotron emission from WR stars, for which there is
now strong observational support \citep{Dougherty:2000b}.  It is
possible that this is also the case in O-type stars with synchrotron
radio emission, \eg \object{Cyg OB2 \#5} \citep{Contreras:1997}, though there
remain several apparently single stars that obviously do not fit a CWB
interpretation.  

To date, modelling has been restricted to the radiometry from such
systems.  At frequency $\nu$, the observed flux ($S_\nu^{\rm obs}$) is
related to the thermal flux ($S_\nu^{\rm th}$), the synchrotron
emission arising from the wind-wind collision region ($S_\nu^{\rm
syn}$), and the free-free opacity ($\tau_\nu^{\rm ff}$) of the
circum-system stellar wind envelope by
\begin{equation}
S_\nu^{\rm obs} = S_\nu^{\rm th} + S_\nu^{\rm syn} e^{-\tau_\nu^{\rm ff}}.
\end{equation} 
It is typically assumed that the stellar wind envelope is radially
symmetric and that the synchrotron emission arises from a point source
at the stagnation point of the two stellar winds~\citep[see][~for
examples]{Williams:1990a,Chapman:1999,Skinner:1999,Monnier:2002}. In
this manner, relatively simple analytical solutions to the radiative
transfer equation can be obtained.  Although these models may
successfully recover the radiometry, a single-valued free-free opacity
determined along the line-of-sight to a point-like synchrotron
emission region is an over-simplification, especially considering the
extended region of synchrotron emission observed from both
\object{WR\,146} and \object{WR\,147}. Furthermore, in a CWB the
assumption of radial symmetry fails in the collision zone.

So far, no attempts have been made to construct synthetic radio images based
on more realistic density and temperature distributions.  This is
surely imperative given the advent of spatially resolved observations
where direct comparison between models and observations may be
expected to increase our understanding of this phenomenon. In making
the first steps toward addressing this situation, we have calculated
the free-free and synchrotron radio emission arising from an early-type
binary system under various simplifying assumptions. Our approach
extends and improves on previous work by including the ability to
simulate both the free-free \emph{and} synchrotron emission and
absorption from the stellar winds and wind-wind collision region based
on 2D, axis-symmetric hydrodynamical simulations of the density and pressure
distribution. Various synchrotron cooling mechanisms can also
be incorporated as required. An arbitrary inclination angle for the 
line of sight can be specified, and the radiative transfer equation is 
then solved to generate synthetic images and spectra.

The layout of the remainder of this paper is as follows. In
Sec.~\ref{sec:modelling} we describe our model in more detail and in
Sec.~\ref{sec:param_study} we perform a parameter study using a
``standard'' model of a very wide CWB. We examine the influence of
synchrotron self-absorption and the Razin effect, which have not
hitherto received much attention, along with system inclination and
binary separation. The application of our model to observations of
\object{WR\,147} is described in Sec.~\ref{sec:147}. In
Sec.~\ref{sec:summary} we summarize and note future directions.  In
Appendix A, the geometry of the ray-tracing technique for solving the
radiative transfer equation is described.

\section{Modelling the radio emission from CWBs}
\label{sec:modelling}
To calculate the radio emission from an early-type binary, we must
first compute the structure of the wind-wind collision. This is most
readily achieved through the use of a hydrodynamical code, and we use
VH-1 \citep{Blondin:1990}, a Lagrangian remap version of the
piecewise-parabolic method~\eg \citet{Pittard:1997}. In this paper we
are concerned particularly with wide systems, so it is assumed that
the spherically symmetric stellar winds are accelerated
instantaneously to terminal speeds. This is reasonable given that the
winds attain terminal velocity at a small percentage of the radial
distance from the stars to the wind collision region.  Distortion of
the collision zone by orbital motion, {most especially close to the
shock apex, is negligible so axis-symmetry is also assumed.}  We use
WR or solar abundances for the stellar winds, as appropriate. Such
models have been successfully applied to the analyses of X-ray
observations from CWBs ~\citep[see][~and references
therein]{Pittard:2002a, Pittard:2002b}.

Once a suitable hydrodynamic solution has been obtained, the necessary
information is read into our radiative transfer ray-tracing code, and
appropriate emission and absorption coefficients for each cell on the
2D grid are determined.  A synthetic image on the plane of the sky is
then generated by solving the radiative transfer equation along
suitable lines of sight through the grid (see Appendix A for more
details). The following sections detail the calculation of the
emission and absorption properties for each hydrodynamic cell.

\subsection{Thermal emission and absorption} 
The thermal emission ($\varepsilon_{\nu}^{\rm ff}$) and absorption
($\alpha_{\nu}^{\rm ff}$) coefficients at a frequency $\nu$ are given
by~\cite{Rybicki:1979} as
\begin{eqnarray}
\varepsilon_{\nu}^{\rm ff}
&=&6.8\times10^{-38}Z^{2}n_{e}n_{i}T^{-1/2}e^{-h\nu/kT}g_{\rm ff},\\
\alpha_{\nu}^{\rm ff}&=&3.7\times10^{8}T^{-1/2}Z^{2}n_{e}n_{i}\nu^{-3}
(1-e^{-h\nu/kT})g_{\rm ff},
\end{eqnarray}
\noindent where $Z$ is the ionic charge, $n_{e}$ and $n_{i}$ are
electron and ion number densities, $T$ is the temperature of the gas,
and $g_{\rm ff}$ is a velocity averaged Gaunt factor
\citep{Hummer:1988}.  The densities $n_{e}$ and $n_{i}$ are determined
from the cell density, composition and ionization, where the
ionization is specified as a function of cell temperature.  The
appropriate ionization of the elemental species for a particular
temperature regime is determined and the absorption and emission
coefficients are then evaluated for several ionic charges.

{The accuracy of the ``thermal'' code was verified by using a simple
spherically symmetric, isothermal stellar wind model with an $r^{-2}$
radial density distribution. The resulting fluxes and free-free
opacities matched those derived from the analytical expressions given
in \citet{Wright:1975}.}

\subsection{Synchrotron emission}
\label{sec:synch}
When non-relativistic particles are accelerated in a magnetic field,
emission occurs at frequencies corresponding to harmonics of the
gyration frequency of the particle. This is known
as cyclotron emission, and most of the radiated power occurs at the
gyration frequency. On the other hand, if the particles are
relativistic, the emission blurs into a continuum rather than a series
of delta functions and emission may occur at a frequency many
orders of magnitude greater than the gyration frequency. The
synchrotron emission from a single electron
is~\citep[\cf][]{Rybicki:1979}

\begin{equation}
\label{eq:sync_single}
P(\nu) = \frac{\sqrt{3} q^{3} B {\rm \;sin\;} \alpha}{m c^{2}} F(\nu/\nu_{c})
\end{equation}

\noindent where $q$ is the particle charge, $B$ is the magnetic field
strength, $\alpha$ is the pitch angle of the particle relative to the
direction of the B-field, $m$ is the mass of the particle, {and
$F(\nu/\nu_c)$ is a dimensionless function describing the total power
spectrum of the synchrotron emission, with $\nu_c$ the frequency where
the spectrum cuts off, given by
\begin{equation}
\nu_c=\frac{3\gamma^2 q B {\rm \;sin\;} \alpha}{4 \pi m c}. 
\end{equation}
Values for $F(\nu/\nu_c)$ are tabulated
in~\citet{Ginzburg:1965}.} Throughout the remainder of this paper we
assume that ${\rm sin}~\alpha = 1$. The emissivity per unit volume
from a mono-energetic distribution of electrons is then simply the
product of Eqn.~\ref{eq:sync_single} and the number density of the
radiating particles.
 
In our model we suppose that the relativistic electrons are created by
1st-order Fermi acceleration at the shocks where the winds collide.
For a strong shock with a compression ratio of 4, test particle theory
predicts that accelerated electrons will have an energy distribution
with a power-law index of 2 \citep[see references
in][]{Eichler:1993}. Hence, the number of relativistic electrons per
unit volume with energies between $\gamma$ and $\gamma + d\gamma$ is
$N(\gamma) d\gamma = C\gamma^{-p}d\gamma$, where $\gamma$ is the
Lorentz factor and $p=2$.  More recent nonlinear calculations yield a
similar energy spectrum for a wide range of shock
parameters \eg~\citet{Ellison:1985}.  A useful discussion of shock
acceleration can be found in~\citet{Ellison:1991}.

For such a power-law distribution of electrons, the total synchrotron
emission at frequency $\nu$ is given by \citep{Rybicki:1979}

\begin{eqnarray}
\label{eq:sync}
P(\nu)=\frac{\sqrt{3}q^{3}CB}{mc^{2}(p+1)} 
\Gamma\left(\frac{p}{4}+\frac{19}{12}\right)
\Gamma\left(\frac{p}{4}-\frac{1}{12}\right)\nonumber \\
\times \left(\frac{2\pi mc\nu}{3qB}\right)^{-(p-1)/2},
\end{eqnarray}

\noindent where $\Gamma$ is the Gamma function.  Eqn.~\ref{eq:sync} is
only strictly valid in the synchrotron limit, when the frequency of
emission is between approximately a few times the gyration frequency
($\nu_B$) and $\nu_{c}$ (defined in Eqn.~\ref{eq:nu_c}), but we
restrict this investigation to frequencies much greater than $\nu_B$
\eg for our model of \object{WR\,147} $\nu_B < 10^4$~Hz at the shock
apex.

To evaluate Eqn.~\ref{eq:sync}, we must determine the appropriate value
for the normalization constant $C$, and the magnetic field strength
$B$. Since our hydrodynamical simulations do not provide direct
information on the magnetic field or relativistic particle
distribution, we follow the standard procedure in such
cases~\citep[\eg][] {Chevalier:1982, Mioduszewski:2001}: that is, to
assume that the magnetic energy density $U_{\rm B}$, and the
relativistic particle energy density $U_{\rm rel}$, are proportional
to the thermal particle internal energy density $U_{\rm th}$. Then

\begin{equation}
U_{\rm B} = B^{2}/ 8\pi = \zeta_{\rm B} U_{\rm th},
\end{equation}

\begin{equation}
U_{\rm rel} = \int n(\gamma) \gamma mc^{2} d\gamma = \zeta_{\rm rel} U_{\rm th},
\end{equation}

\noindent where $\zeta_{\rm B}$ and $\zeta_{\rm rel}$ are constants
which are typically set to around 1\%~\citep{Mioduszewski:2001},
and $U_{\rm th}={3\over2}P$
where $P$ is the gas pressure and the adiabatic index of $5/3$ is
assumed, as for an ideal gas. The common assumption of equipartition
corresponds to $\zeta_{\rm B} = \zeta_{\rm rel} = \zeta$. 

As noted in \citet{Mioduszewski:2001}, there is some justification for
such a relation between the energy densities: the thermal energy
density is higher in the post-shock than in the pre-shock gas, and
this is also where the magnetic field is likely to be enhanced, and
the particles accelerated to relativistic energies.  However, such
arguments ultimately need to be replaced by detailed physical
processes. It is well established that the shock acceleration of
protons is very efficient, but how much energy is transferred to
electrons in shocks is much less certain. We note that the
relativistic electron energy is unlikely to be more than 5\% of the
total available shock energy, and could conceivably be much
less~\citep{Blandford:1987, Ellison:1991,Eichler:1993}.

For $p=2$,

\begin{equation}
\label{eq:c}
C = \frac{U_{\rm rel}}{m_{e} c^{2} {\rm ln} \gamma_{\rm max}}, 
\end{equation}

\noindent where we have assumed $1<\gamma<\gamma_{\rm max}$ and
$\gamma_{\rm max}$ specifies the maximum energy of the relativistic
electrons. Since it is assumed in Eqn.~\ref{eq:sync} that $\gamma_{\rm
max} \rightarrow \infty$, the synchrotron emission extends to
infinitely high frequencies with a spectrum $\propto \nu^{-0.5}$.  In
reality the synchrotron flux will depart from this relationship at
frequencies near $\nu_{c}$, where $\nu_{c}$ is the frequency of peak
emission from a particle of energy $E = \gamma_{\rm max} mc^{2}$,
given by \citep{Rybicki:1979}

\begin{equation}
\label{eq:nu_c}
\nu_{c} \approx \frac{3}{4 \pi}\frac{ \gamma_{\rm max}^{2} q B} {m c}.
\end{equation}

\noindent In CWBs we expect $\gamma_{\rm max}$ to
be set by the balance between energy gain by Fermi acceleration and
various cooling processes, which include inverse-Compton (hereafter
IC) cooling, synchrotron decay, and ion-neutral wave damping. Ion-neutral
damping does not play a dominant role in CWBs~\citep{Eichler:1993}, and is
not discussed further here. The relevant processes and timescales are
discussed in the following subsections.

\subsection{Synchrotron self-absorption}
We include synchrotron self-absorption (hereafter SSA) in our models.
For a power-law distribution of particle energies, the absorption
coefficient for SSA is~\citep{Rybicki:1979}
\begin{eqnarray}
\label{eq:sync_abs}
\alpha_{\nu} = \frac{\sqrt{3} q^{3}}{8\pi m}
\left(\frac{3q}{2\pi m^{3}c^{5}}\right)^{p/2}
C B^{(p+2)/2} \nonumber \\
\Gamma\left(\frac{3p+12}{12}\right)\Gamma\left(
\frac{3p+22}{12}\right) \nu^{-(p+4)/2}, 
\end{eqnarray}
where the symbols have their previously defined meanings.

\subsection{The Razin effect}
When relativistic charges are surrounded by a plasma (as opposed to
existing in a vacuum), the beaming effect that characterizes synchrotron
radiation is suppressed. In essence, the refractive index of the medium 
reduces the Lorentz factor of the charge to

\begin{equation}
\gamma' = \frac{\gamma}{\sqrt{1 + \gamma^{2}\nu_{0}^{2}/\nu^{2}}},
\end{equation}
 
\noindent where the plasma frequency $\nu_{0} = \sqrt{q^{2}n/\pi m}$.
Hence synchrotron emission by relativistic charges in the medium is
possible only when $\gamma' >> 1$, and a low frequency cut-off occurs. 
Quantitative details of this process, which is known as the 
Tsytovitch-Eidman-Razin effect, can be found in \cite{Hornby:1966}. 
The characteristic cut-off frequency is given by

\begin{equation}
\nu_{R} = 20 \frac{n_{e}}{B}.
\label{eq:razin_freq}
\end{equation}

\noindent Since the emission decreases exponentially at frequencies
low compared with $\nu_{R}$, and there is a noticeable effect at
frequencies greater than $\nu_{R}$, \eg there is a 10\% reduction in
flux at $\nu = 10\nu_{R}$, we approximate this reduction by
multiplying the synchrotron flux from our model by a factor
$e^{-\nu_{R}/\nu}$.

\subsection{The low and high-energy cutoff in the relativistic particle distribution}
\label{sec:cutoffs}
The nature of the synchrotron spectrum is determined by the underlying
relativistic particle energy distribution. The low and high energy
cutoffs in the energy distribution are particularly important.  At low
energies, the relativistic particle distribution can be potentially
affected by coulombic collisions with thermal ions. However, at the
densities and temperatures in the wind-collision regions of the wide
systems considered in this paper this process is insignificant.  The
high energy cutoff is determined at the point where the rate of energy
gain balances the rate of energy loss. Radiation losses due to
synchrotron emission and IC scattering are in the same ratio as the
magnetic field energy density to the photon energy density,

\begin{equation}
\frac{P_{\rm sync}}{P_{\rm compt}} = \frac{U_{\rm B}}{U_{\rm ph}}.
\end{equation}

\noindent As noted in Sec.~\ref{sec:synch}, it is expected that
$\zeta_{\rm B}$ is no more than 0.05, which suggests that $P_{\rm
sync} < P_{\rm compt}$ in the wide CWB systems under consideration,
and IC cooling dominates over synchrotron losses.

Electrons are rapidly accelerated to relativistic energies at the
shocks bounding the wind-collision zone, and attain an energy spectrum
specified by $p=2$ and $\gamma_{\rm max}$ (where the rate of energy
loss by IC cooling balances the rate of energy gain from 1st order
Fermi acceleration). They are then advected out of the shock and (for
$\gamma < 10^{5}$) are effectively frozen into the post-shock
flow~\citep{White:1985}. If the characteristic flow time out of the
system is shorter than the timescale for IC cooling, the dominant
cooling process is adiabatic expansion. This is the case for wide CWBs
and low $\gamma$, and since adiabatic cooling is treated by the
hydrodynamic code, we assume that $\zeta_{\rm B}$ and $\zeta_{\rm
rel}$ are spatially invariant in the post-shock flow. For high
$\gamma$, IC losses can be very rapid even in the widest systems, as
we show is the case for \object{WR\,147} in Sec.~~\ref{sec:147}. IC cooling will be an
even more important consideration in closer CWBs \eg \object{WR\,140} and
shorter period systems, where it could {\em prevent} the acceleration
of electrons to even fairly modest values of $\gamma$. In these first
attempts at modelling the radio emission from wide CWBs we have not
included IC cooling explicitly in the code, but where relevant we
discuss its impact on the resulting spectra \eg
Secs.~\ref{sec:binary_sep} and~\ref{sec:147}.  This will be
treated more fully in Pittard \etal (in preparation).

\subsection{Some simplifying assumptions}
For simplicity in this first investigation, we assume that the
post-shock ion and electron temperatures equalize very rapidly. While
the equilibration timescale can be significant compared with the flow
timescale \eg \object{WR\,140} - see ~\citet{Usov:1992,Zhekov:2000}, the exact
situation is currently unclear with the importance of possible
electron heating mechanisms in collision-less shocks still being
debated.  We anticipate that our simplification will have little
bearing on the resulting radio emission because the synchrotron
emission from the wind-collision zone dominates the thermal free-free
emission. While it may affect the efficiency with which the
accelerated electrons are extracted from the thermal pool, this
conjecture is impossible to test currently. Instead its effect is
simply folded into the value of $\zeta$. Also, we assume that the
gas is in ionization equilibrium, which is not most likely the case
immediately post-shock, especially in wide systems. However, we
believe that this will not impact the results presented here since the
bulk of the emission from the shocked gas is synchrotron, not
free-free emission.

One final assumption in our models is that the magnetic field in the
post-shock gas is highly tangled. Conveniently, this allows us to treat
the synchrotron emission and absorption as isotropic, and means that
it is possible to use isotropic formulae to calculate the IC losses
(Pittard \etal, in preparation). While the magnetic
field from each star should be structured on large scales, and
toroidal in shape at the large distances considered in the models
presented here~\citep[see][and references therein]{Eichler:1993}, one
might expect finer-scale structure in the post-shock flow. However,
such details are beyond the scope of this paper.

\section{Parameter Study}
\label{sec:param_study}

\begin{figure}[t]
\vspace{6.15cm}
\includegraphics{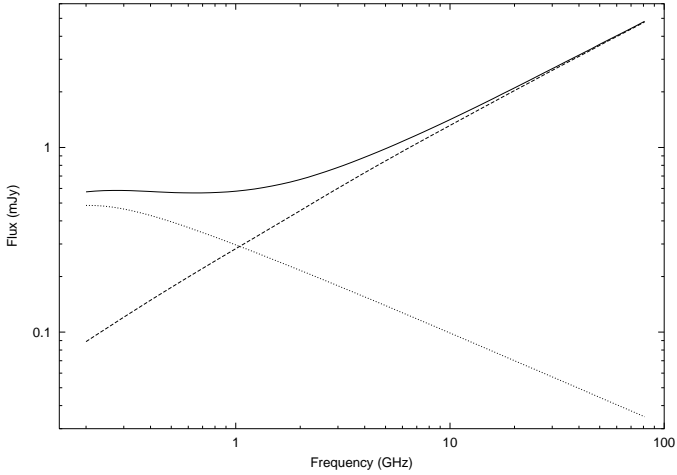}
\caption[]{Spectra from our standard model with $0^\circ$ inclination
  - free-free flux (dashed), synchrotron flux (dotted), and the total
  flux (solid). The effect of free-free absorption on the synchrotron
  emission in this wide system is only just evident below 300 MHz.
  The spectral index of the thermal spectrum is $+0.6$, and the
  optically thin part of the synchrotron spectrum is $-0.5$. Neither
  the Razin effect nor SSA are treated in this model. }
\label{fig:standard_spec}
\end{figure}

\subsection{A standard model}
\label{sec:standard_model}
Our initial investigations have been based around a standard CWB model
with the following parameters: $\Mdot_{\rm WR} = 2 \times
10^{-5}\;\Msolpyr$, $\Mdot_{\rm OB} = 2 \times 10^{-6}\;\Msolpyr$,
$v_{\infty,{\rm WR}} = v_{\infty,{\rm OB}} = 2000\;\kmps$, $D = 2
\times 10^{15}\;\cm$, and $T_{\rm wind}$ for both stars of
20~kK. These wind values are typical of WR and OB stars.  At a
frequency $\nu = 5\;{\rm GHz}$, the adopted binary separation $D$, is
approximately $10\times$ the radius of the $\tau_{\rm ff}=1$ surface
of the WR wind. For inclination angles $\sim 0^\circ$, the lines of
sight to the wind-collision zone are then optically thin, permitting
investigation of the resulting emission from the wind-collision region
in the absence of strong free-free absorption from the circum-binary
stellar wind envelope. We also assume solar abundances for the OB star
and mass fractions $X=0,Y=0.75,Z=0.25$ appropriate for a WC-type star,
{and an ionization structure of H$^+$, He$^+$ and CNO$^{2+}$}. We adopt
$\zeta=10^{-4}$ as our ``canonical'' value of $U_{\rm B}/U_{\rm th}$
and place our model system at a distance of 1.0 kpc. Maximum values of
temperature, density, and magnetic field occur at the shock apex and
for these input parameters we find $T_{\rm max} = 1.75 \times
10^{8}$~K, $\rho_{\rm max} = 9 \times 10^{-19} \;{\rm g\,cm^{-3}}$
($n_{\rm max} = 4 \times 10^{5}\;{\rm cm^{-3}}$), and $B_{\rm max} =
6\;{\rm mG}$.  Though $B$ is an equipartition value, fields of this
order at the location of the shock extrapolate to surface fields on
the OB star of $\sim$ 30-300~G, using the magnetic field structure
given in~\citet{Eichler:1993} and {taking the typical rotation
speed of O-type stars to be $\sim0.1v_{\infty,{\rm
OB}}$~\citep{Conti:1977}.} Such surface fields are consistent with the
few observations of the fields in these types of
stars~\citep{Donati:2001, Donati:2002}.  For all of the models
presented in this paper we fix $\gamma_{\rm max} = 4000$. Since we are
using Eqn.~\ref{eq:sync} to calculate the synchrotron luminosity,
fixing $\gamma_{\rm max}$ only has a weak impact on the constant
$C\sim1/{\rm ln}~\gamma_{\rm max}$.

\begin{figure}[t]
\vspace{15.2cm}
\includegraphics{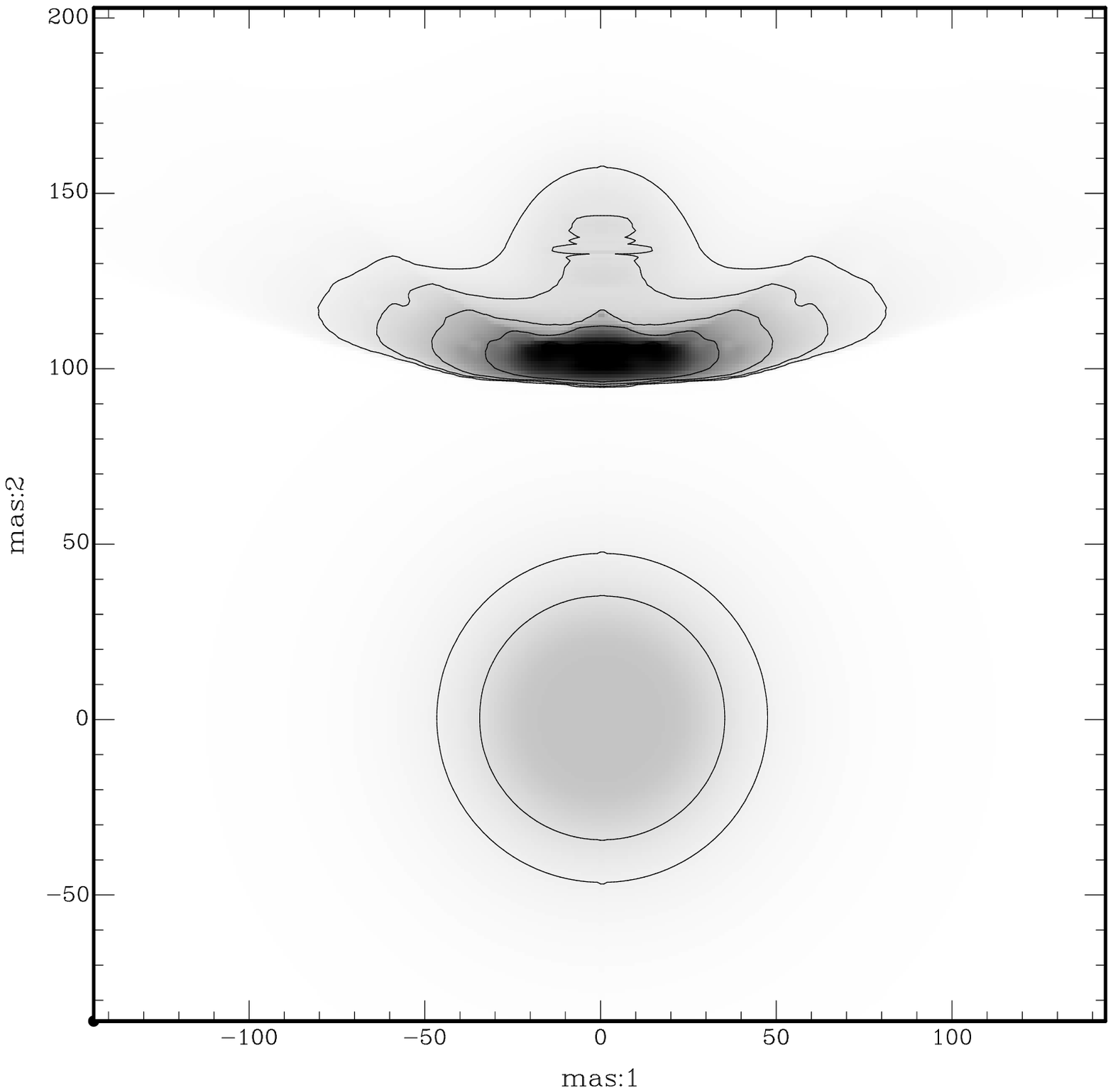}
\includegraphics{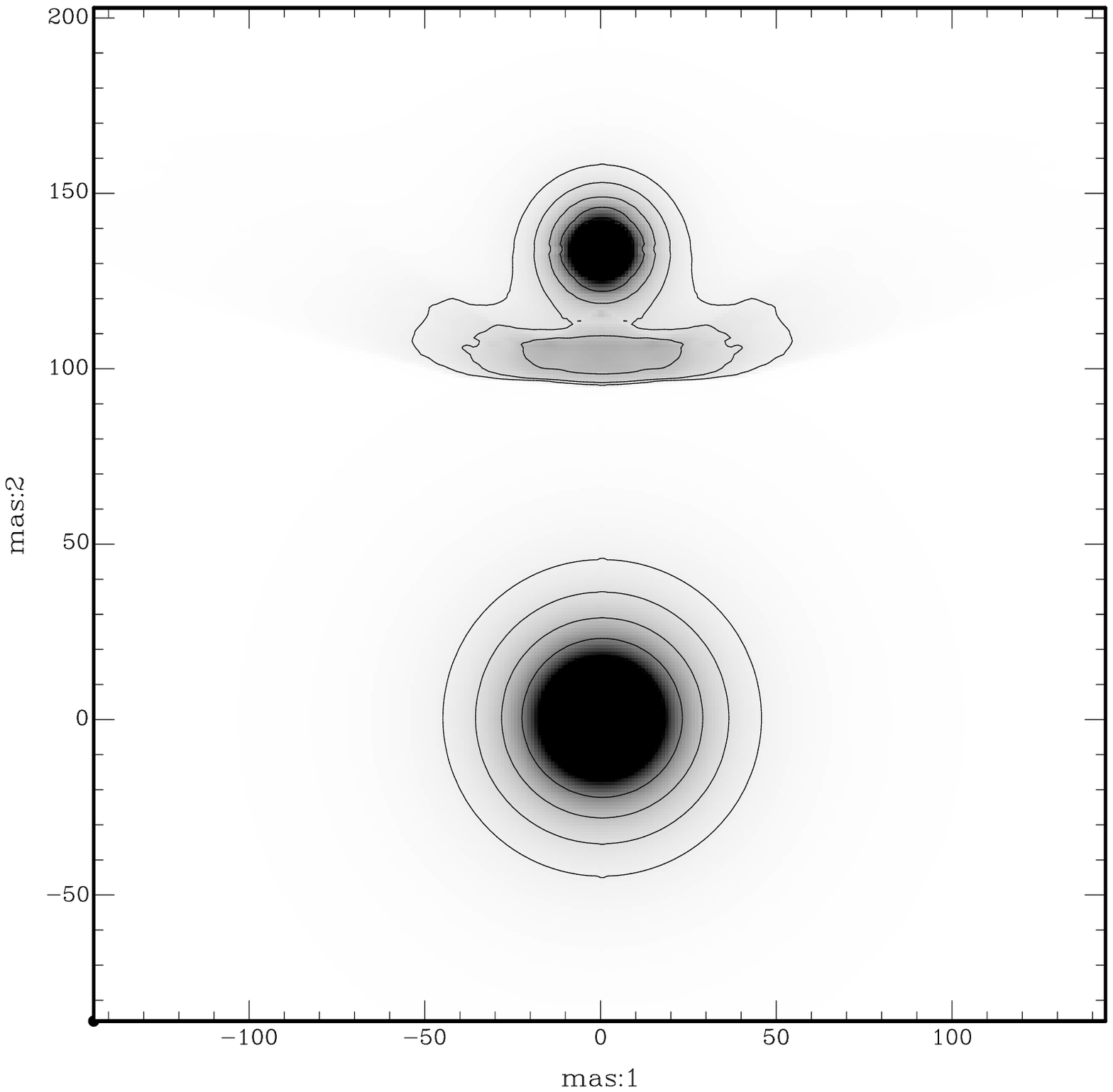}
\caption[]{Intensity distributions of our standard model at $0^\circ$
inclination at 1.6~GHz (top) and 22~GHz (bottom).  Both images have
the same intensity scale, highlighting the relative importance of the
emission from the wind-wind collision region and the stellar winds as
frequency varies (see also Fig.~\ref{fig:147_model_images}). Though
not shown, the emission extends to the panel boundaries. Neither the 
Razin effect nor SSA are included in this calculation.}
\label{fig:standard_distrib}
\end{figure}

The radiometry and intensity distributions obtained from our standard
model are shown in Figs.~\ref{fig:standard_spec} and
~\ref{fig:standard_distrib}. In these first examples we only consider
the impact of free-free absorption: SSA and the Razin effect are
examined in the next section.  Around 1~GHz the synchrotron emission
from the wind-wind collision is comparable to the total free-free
emission, while at 22~GHz the emission is largely from the two stellar
winds. Below 1~GHz, the synchrotron emission dominates the total flux
and the beginnings of a turnover seen at $\sim 300$~MHz in the
synchrotron spectrum is due to free-free absorption from the unshocked
stellar winds. Above 300~MHz, the synchrotron spectrum is a power-law
with a spectral index equal to $(1-p)/2=-0.5$ for $p=2$. The
thermal spectrum is also a power-law with a spectral index of $+0.6$,
as expected for a fully ionised, isothermal stellar wind envelope with
a $r^{-2}$ radial density distribution.

Although we have deliberately avoided modelling a specific system in
this first analysis, these initial results already bear similarities
to observations \eg \object{WR\,147} - see Sec.~\ref{sec:147}. The total flux,
the spectral shape, and the spatial intensity distribution are of the
correct order of magnitude and morphology.

\begin{figure}
\vspace{6.15cm} 
\includegraphics{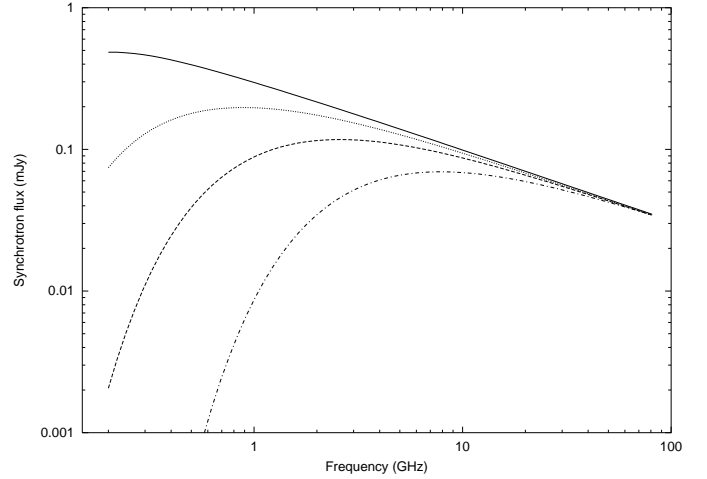}
\caption[]{The influence of the Razin effect on synthetic synchrotron
spectra using the standard model at $0^\circ$ inclination for various
values of $\zeta$. The effect of free-free absorption by the
circum-binary stellar wind envelope is negligible, just visible at
around 200-300 MHz.  No Razin (solid), $\zeta=10^{-3}$ (dotted),
$10^{-4}$ (dashed),$10^{-5}$ (dash-dotted).  All spectra are
normalized to the 80 GHz synchrotron flux when no Razin effect is
included.}
\label{fig:razin_frac_spec}
\end{figure}

\begin{figure}
\vspace{6.15cm}
\includegraphics{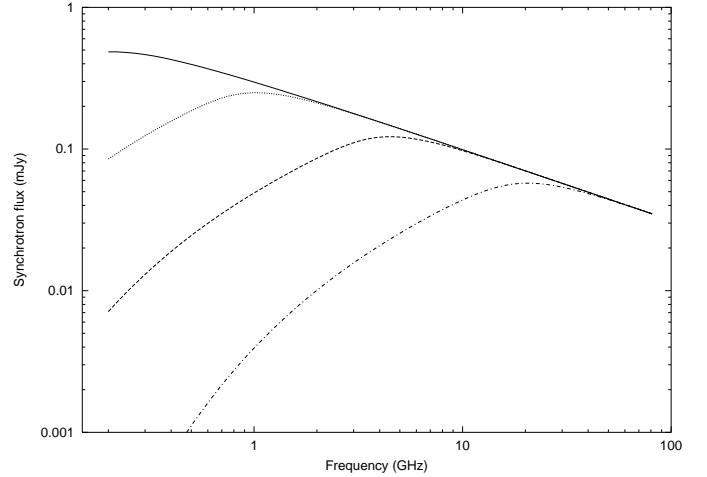}
\caption[]{The effect of SSA on synthetic synchrotron spectra using
the standard model at $0^\circ$ inclination for various values of
$\zeta$. The effect of free-free absorption by the circum-binary
stellar wind envelope is negligible, just visible at around 200-300
MHz.  No SSA (solid), $\zeta=10^{-4}$ (dotted), $10^{-3}$
(dashed),$10^{-2}$ (dash-dotted).  All spectra are normalized to the
80 GHz synchrotron flux when no SSA is included.}
\label{fig:ssa_frac_spec}
\end{figure}

\subsection{Razin effect and Synchrotron Self-Absorption}
\label{sec:raz_ssa}
In Figs.~\ref{fig:razin_frac_spec} and~\ref{fig:ssa_frac_spec} the
influence of the Razin effect and SSA on the synchrotron spectrum of
our standard model is shown compared with the case where neither
mechanism is active. Note that in the models shown, free-free opacity
is negligible and has little influence on the spectra. It is clear
that both the Razin effect and SSA can be important in quenching the
low-frequency spectrum, with the relative impact of these two
mechanisms in these models being strongly dependent on the value of
$\zeta$.  For $\zeta=10^{-4}$, the Razin effect is the dominant
mechanism whereas for $\zeta=10^{-3}$, SSA dominates. From
Eqs.~\ref{eq:sync_abs} and~\ref{eq:razin_freq}, $\alpha_{\nu}\propto
\zeta^2$ and $\nu_{R}\propto \zeta^{-{1\over2}}$ (for fixed $n_{e}$)
so we expect SSA to be increasingly important as $\zeta$ increases,
with the Razin effect becoming more important for decreasing
$\zeta$. For our standard model and reasonable values of $\gamma_{\rm
max}$, we find that the Razin effect dominates the low-frequency
absorption when $\zeta \ltsimm 10^{-4}$, the Razin effect and SSA are
roughly comparable when $\zeta \sim 10^{-3}$, and SSA dominates when
$\zeta \gtsimm 10^{-2}$.  Of course, we expect these values to be
different for other model systems.

For a given value of $\zeta$, a different value of $\gamma_{\rm max}$
would affect the value of $C$ (Eqn.~\ref{eq:c}) and hence the intrinsic
synchrotron flux, $P(\nu)$ (Eqn.~\ref{eq:sync}), although the
synchrotron spectrum would still extend to infinite frequency as a
result of adopting Eqn.~\ref{eq:sync}. To obtain a certain intrinsic
flux, only the combination $\zeta^{7/4}/{\rm ln} \gamma_{\rm max}$ is
required to be a specific value and $\zeta$ and $\gamma_{\rm max}$ can
vary within this constraint.  However, since $\alpha_{\nu}$
(Eqn.~\ref{eq:sync_abs}) and $\nu_{R}$ (Eqn.~\ref{eq:razin_freq}) depend
on different powers of $\zeta$ and $\gamma_{\rm max}$, this degeneracy
breaks down when SSA and the Razin effect are important. This means
that the application of our model is dependent on the particular value
of $\gamma_{\rm max}$ assumed. Nevertheless, since ${\rm ln}
\gamma_{\rm max}$ will vary by less than a factor of 4 for a
reasonable range of $\gamma_{\rm max}$, the strength of these
absorption terms will not change a great deal.  In fact, as
$\alpha_{\nu} \propto \zeta^{2}/{\rm ln} \gamma_{\rm max}$ \ie almost
the same dependence as the intrinsic flux, its variation will be very
small. As $\nu_{R} \propto \zeta^{-1/2}$, the relative strength of the
Razin effect is slightly more sensitive to our particular choice of
$\gamma_{\rm max}$, but not hugely so. Hence the curves in
Figs.~\ref{fig:razin_frac_spec} and~\ref{fig:ssa_frac_spec} are not
strictly functions of only $\zeta$, as there is a slight implicit
dependence on $\gamma_{\rm max}$ as well.

Figs.~\ref{fig:razin_frac_spec} and~\ref{fig:ssa_frac_spec} also
demonstrate that the spectral shape belies the underlying
mechanism; the SSA spectra are power-law, whereas the Razin dominated
spectra decay exponentially.  Optically thick synchrotron emission
actually has a power-law spectrum with a slope of +2.5, but we find
that the spectral slope in our calculations is $\sim +1$. This
difference reflects the fact that the synchrotron emission in our
models is always a mixture of optically thick and thin emission. When
the synchrotron emission from the shock apex is optically thick, a
substantial amount of optically thin emission from the downstream flow
contributes to the total non-thermal emission.

\subsection{Effect of Binary Separation}
\label{sec:binary_sep}

The dramatic variations in the radio flux of the highly eccentric
WR+OB binary \object{WR\,140} point to the significance of binary separation to
the observed radio emission. In \object{WR\,140}, the separation varies from
$\sim2$~AU at periastron to $\sim28$~AU at apastron, a separation
range over which different physical mechanisms determine the
resulting synchrotron spectrum.  In this section, we perform
calculations with varying binary separation to investigate the effect
on the intrinsic synchrotron luminosity, and also how separation
affects the relative importance of free-free opacity, the Razin
effect, and SSA on the spectrum. We also discuss the impact of IC
cooling as a function of separation.
\begin{figure}[t]
\vspace{6.15cm}
\includegraphics{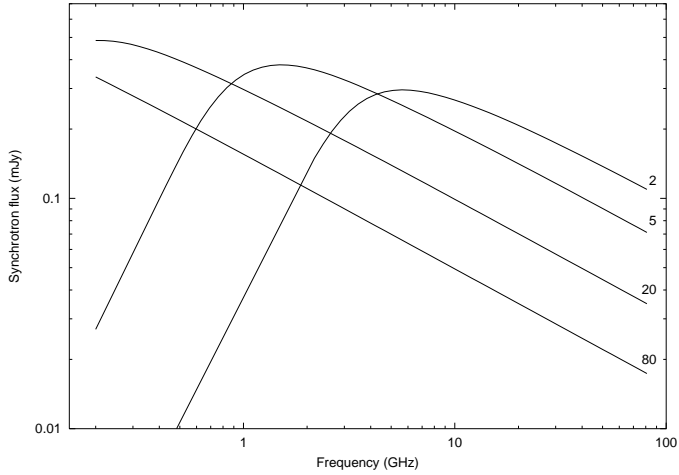}
\caption[]{The effect of separation on synchrotron spectra using the
standard model at $0^\circ$ inclination and $\zeta=10^{-4}$, for
$D=2, 5, 20$, and $80\times 10^{14}$ cm. Only the effect of free-free opacity
is included.}
\label{fig:sep_ffspec}
\end{figure}

\begin{figure}
\vspace{6.15cm}
\includegraphics{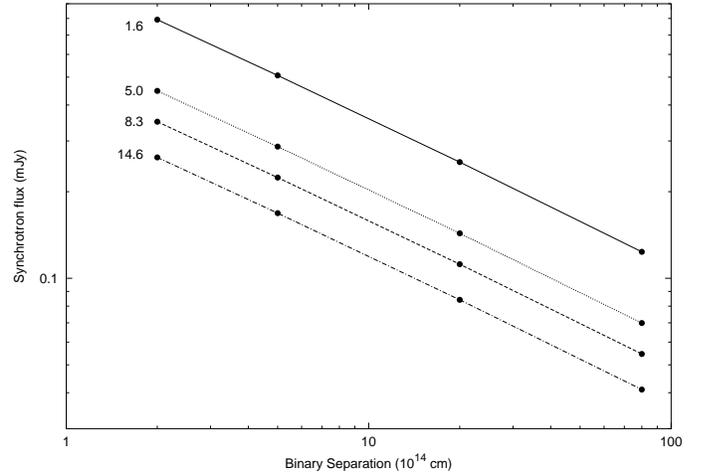}
\caption[]{The effect of separation on the synchrotron luminosity for
a power-law electron energy distribution in the optically thin
regime. Shown are the data points at four frequencies: 1.6 (solid), 5.0
(dotted), 8.3 (dashed), 14.6~GHz (dot-dashed).  The slope of this
log-log plot is -1/2, as expected.}
\label{fig:sync_lum_dist}
\end{figure}
\begin{figure}[t]
\vspace{6.15cm}
\includegraphics{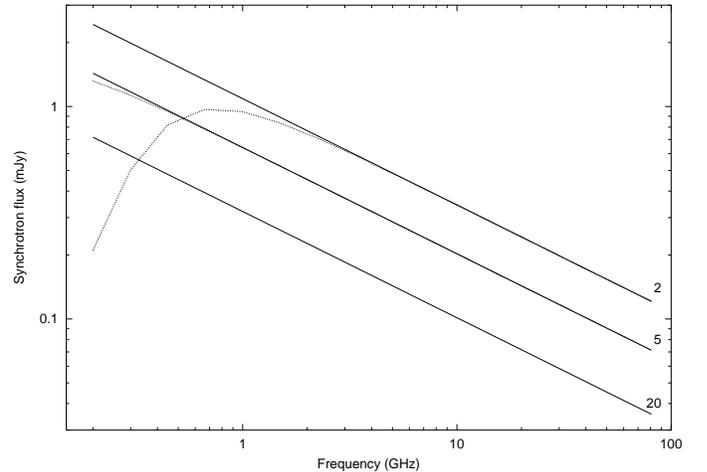}
\caption[]{The effect of separation on the free-free opacity {\em
within} the wind-wind collision region, using the standard model at
$0^\circ$ inclination and $\zeta=10^{-4}$, for $D=2, 5$, and 
$20\times 10^{14}$ cm. The intrinsic synchrotron emission is shown (solid)
along with the spectra of the synchrotron emission emerging from
the collision zone (dotted). }
\label{fig:sep_wwcffspec}
\end{figure}
\begin{figure}
\vspace{6.15cm}
\includegraphics{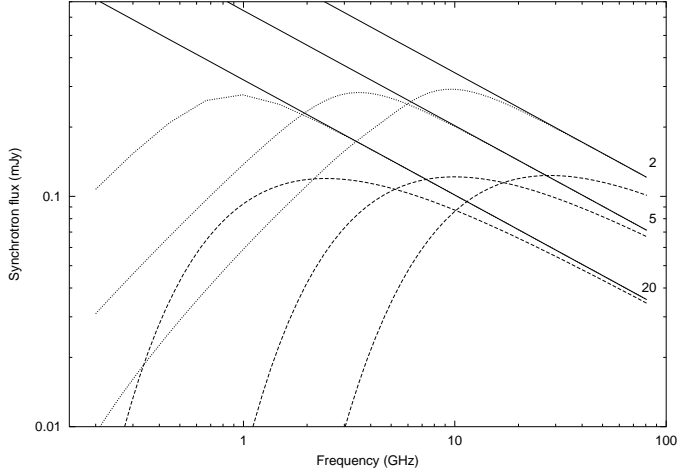}
\caption[]{The effect of separation on the synchrotron emission using
the standard model at $0^\circ$ inclination and $\zeta=10^{-4}$, for
$D=20\times 10^{14}$ (labelled ``20''), $5\times 10^{14}$ cm (``5'')
and $D=2\times 10^{14}$ (``2''): the intrinsic synchrotron
emission \ie no Razin or SSA (solid); only Razin effect included (dashed);
only SSA included (dotted). The intrinsic synchrotron emission is shown to
highlight the effect of separation on the synchrotron emission \ie in
the absence of the strong free-free absorption, as seen in
Fig.~\ref{fig:sep_ffspec}.}
\label{fig:sep_intrin_spec}
\end{figure}

In Fig.~\ref{fig:sep_ffspec} we show the effect of free-free opacity
on the spectra resulting from varying the separation between $2 \times
10^{14}$~cm and $80 \times 10^{14} \cm$, while the other parameters of
our standard model were unchanged. The radius of optical depth unity
for the WR wind is $\approx 2 \times 10^{14} \cm$ at 5~GHz. The
increasing importance of free-free absorption as the separation
decreases is clearly observed at the lower frequencies. As the stars
move closer together the wind-wind-collision region moves closer to
both the WR star and the companion, and is surrounded by increasingly
dense gas which increases the line-of-sight opacity to the shock
apex. For a stellar wind with an $r^{-2}$ radial density distribution
it can be easily shown that the line-of-sight opacity $\tau$, at frequency
$\nu$ through the wind is $\tau\propto \xi^{-3}\nu^{-2.1}$, where $\xi$
is the impact parameter.  Since $\xi \propto D$ for a given
inclination, the turnover frequency for a constant opacity value is
$\nu\propto D^{-10/7}$, in excellent agreement with the spectra in
Fig.~\ref{fig:sep_ffspec}.

With the parameters for our standard model, Fig.~\ref{fig:sep_ffspec}
also reveals that the synchrotron luminosity increases as the separation
decreases, due to the increased thermal energy density in the
collision region. From Eqn.~\ref{eq:sync}, it can be deduced that the
intrinsic synchrotron emission per unit volume $P(\nu) \propto
\zeta^{3/4}n^{3/4}\nu^{-1/2}$ for an electron power-law index $p=2$,
in the absence of SSA or the Razin effect.  Since the post-shock
density $\propto D^{-2}$ and the volume of the emitting region scales
as $D^3$, the total synchrotron emission from the entire wind
collision volume $\propto D^{-1/2}\nu^{-1/2}$. This is illustrated in
Fig.~\ref{fig:sync_lum_dist}. For comparison, we note that the total
intrinsic X-ray emission from the wind-wind collision scales as
$D^{-1}$ in the optically thin,  adiabatic limit \citep{Stevens:1992}.

\begin{figure}
\vspace{17.8cm}
\includegraphics{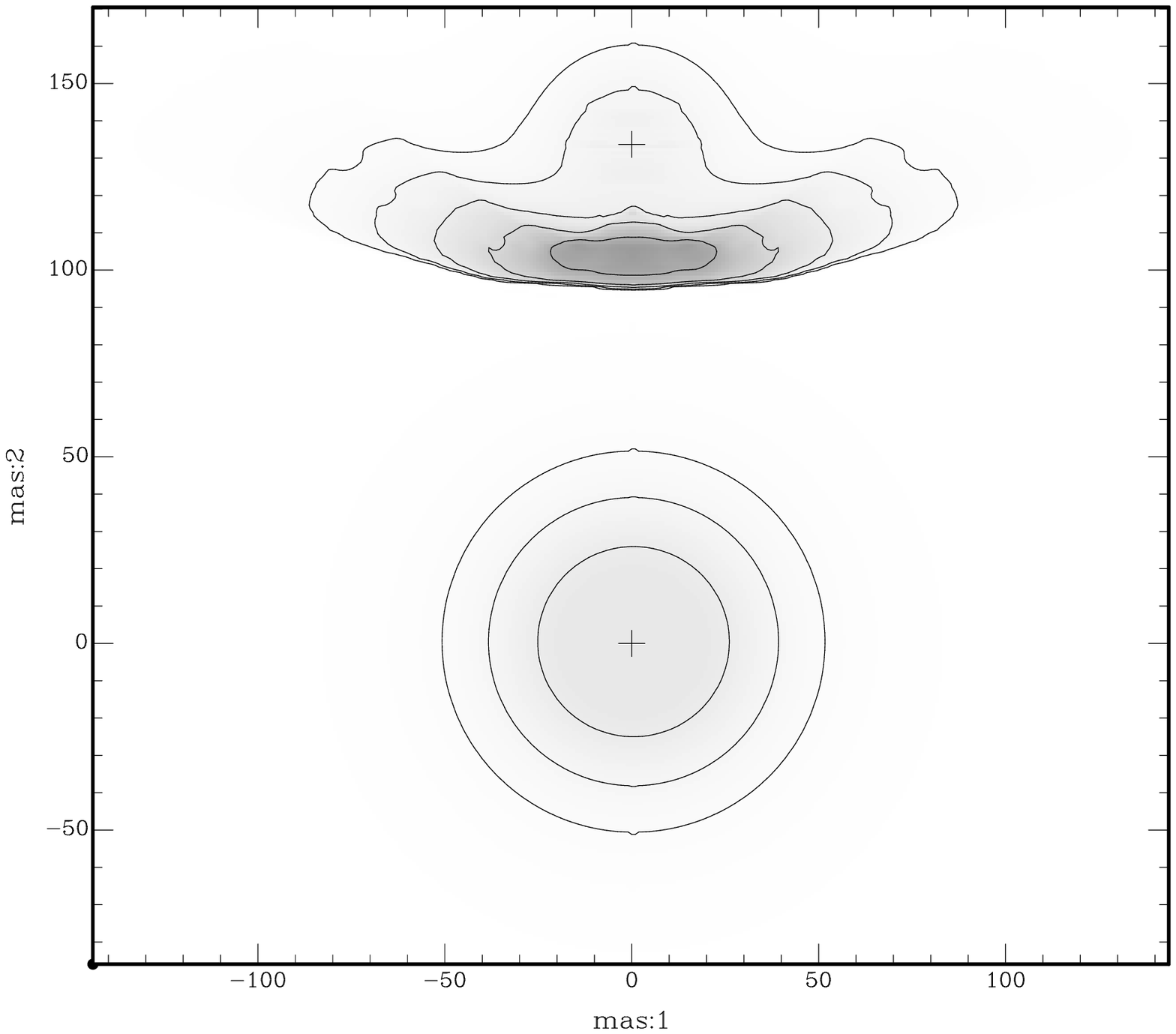}
\includegraphics{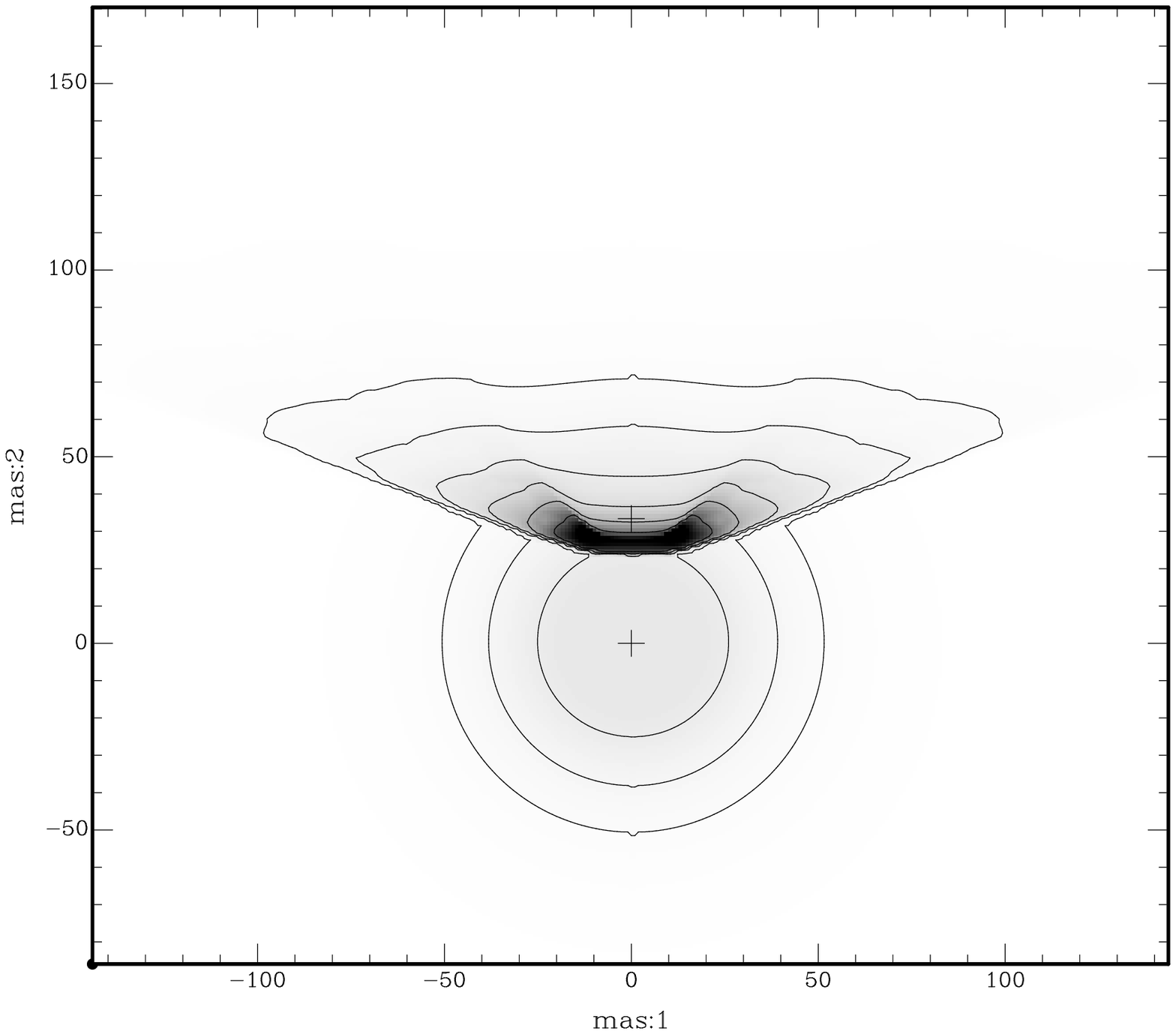}
\includegraphics{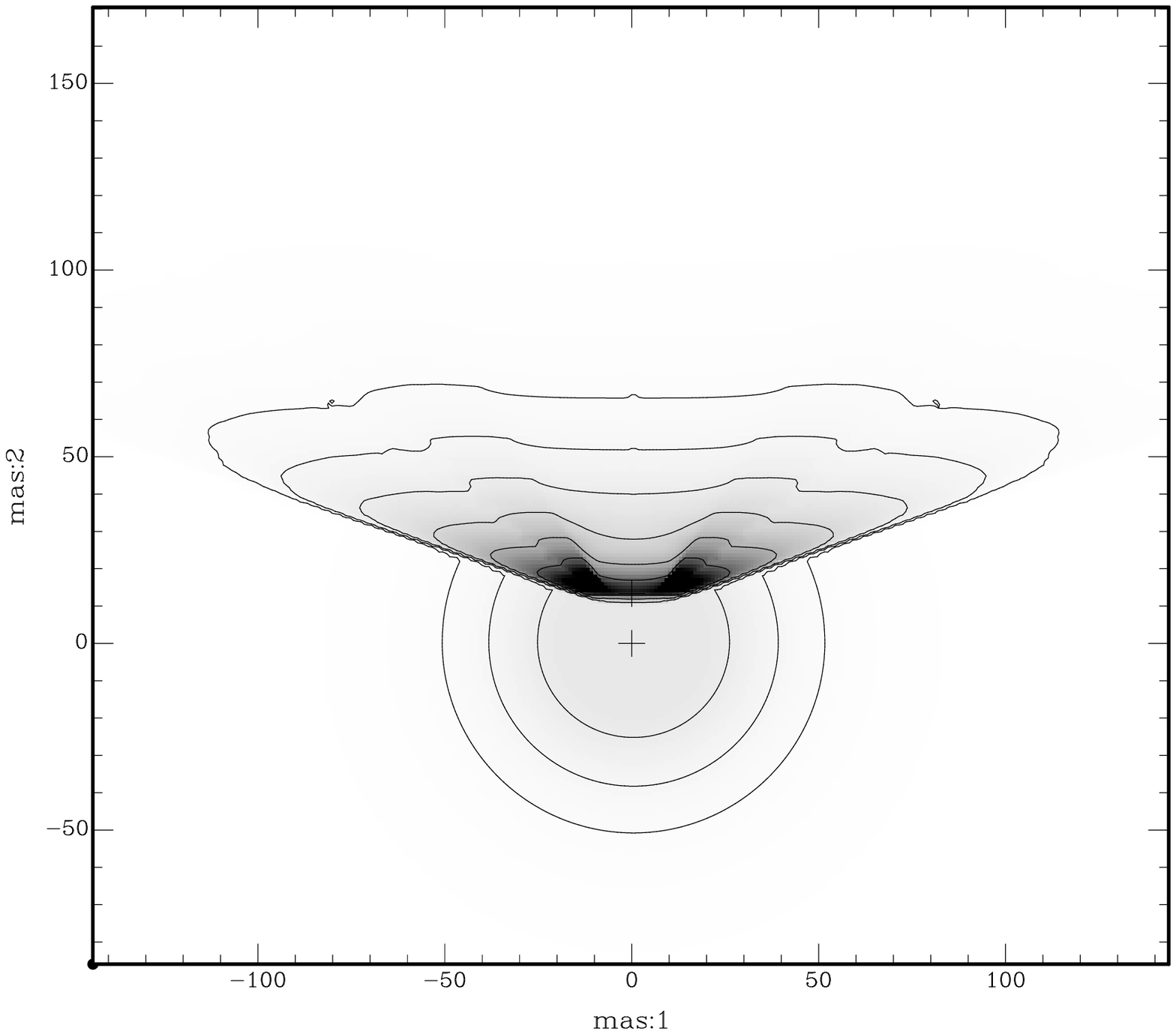}
\caption[]{Intensity distributions at 1.6~GHz and $0^{\circ}$
  inclination for an orbital separation of $20$ (top), $5$ (middle),
  and $2 \times 10^{14}$~cm (bottom), corresponding to angular
  separations of 133, 33 and 13.3~mas respectively. The crosses denote
  the positions of the stars, with the WR star located at (0,0). Each
  image has the same intensity scale and contours. The contours were
  chosen to highlight the relative brightness of the stellar wind of
  the WR star and the emission from the collision region. Though not
  shown, the emission extends to the boundaries of the panels (see
  also Fig.~\ref{fig:147_model_images}). }
\label{fig:dsep_com}
\end{figure}

In Fig.~\ref{fig:sep_wwcffspec} the effect of free-free opacity from
within the shocked gas is shown. As the binary separation is reduced
the free-free opacity in the shocked region becomes increasingly
important, as density increases. Note that shocked gas temperature is
independent of separation. As before, the free-free spectral turnover
$\propto D^{-10/7}$, in close agreement with the spectra shown. In
this case, in systems such as \object{WR\,140} with periaston separation
$\sim3\times10^{13}$~cm, the free-free opacity of the shocked plasma
will have an impact of the spectrum below $\sim10$~GHz.

The impact of changing separation on the importance of the Razin
effect and SSA is shown in Fig.~\ref{fig:sep_intrin_spec}.  The
intrinsic synchrotron spectrum is shown in order that the dramatic
effects of the free-free opacity of both the unshocked circum-binary
stellar winds and the shocked gas, clearly seen in
Fig.~\ref{fig:sep_ffspec} and ~\ref{fig:sep_wwcffspec}, are removed.
As expected, the attenuation from both processes is more pronounced as
the separation decreases since the shock apex occurs in a higher
density part of the stellar wind envelope. The Razin turnover
frequency $\nu_{R} \propto n/B$, and as $B\propto n^{1\over 2}$ (for
fixed $\zeta$), $\nu_R$ increases as $n^{1\over2}$. Because $n \propto
D^{-2}$, we find that $\nu_{R} \propto D^{-1}$. With $D=2\times
10^{15}$~cm, our standard model gives $\nu_{R} = 1.3$~GHz for emission
at the shock apex. Away from the shock apex $\nu_{R}$ will be
smaller.  For our model with $D = 5 \times 10^{14} \cm$, $\nu_{R} =
5$~GHz. Both of these values are in excellent agreement with the
frequency where the Razin effect causes a $1/e$ reduction in flux, as
seen in Fig.~\ref{fig:sep_ffspec}.  The opacity due to SSA is
$\tau_{\rm SSA}\sim\alpha_{\nu}r_{\rm OB}$, where $r_{\rm OB}$ is the
distance between the shock apex and the OB star.  Since
$\alpha_{\nu}\propto D^{-4}\nu^{-3}$ and $r_{\rm OB}\propto D$ (see
Eqn.~\ref{eq:rob}), this leads to $\nu_{\rm SSA}\propto D^{-1}$, which
is in excellent agreement with the spectra shown in
Fig.~\ref{fig:sep_ffspec}, where the SSA turnovers occur at
approximately $0.9, 3.5$, and $9.0$~GHz for separations of $2, 5$ and
$20\times10^{14}$~cm respectively.

Also of interest is how the distribution of the synchrotron emission
is affected by changing separation. This is shown in
Fig.~\ref{fig:dsep_com} for three different separations.  In the
middle and bottom images, the maximum intensity actually occurs to
either side of the axis of symmetry, this offset widening as the
separation decreases. This is consistent with significant absorption
along the line of sight to the shock apex, and slightly reduced
absorption to the positions of peak emission. In addition, a given
flux intensity from the wind collision is apparent out to greater
off-axis distances as the separation decreases. This is due to a
combination of increased energy density in the collision zone and the
importance of absorption from the surrounding winds.

As mentioned in Sec.~\ref{sec:cutoffs}, IC scattering is an important
cooling mechanism for relativistic electrons.  To estimate when IC
cooling needs to be considered, we determine the minimum value of
$\gamma$ at which IC cooling can have an effect before the
relativistic electrons adiabatically flow out of the system.  The
distance from the OB companion to the shock apex is given by
\begin{equation}
r_{\rm OB}={\eta^{1\over2}\over{{1+\eta^{1\over2}}}} D.
\label{eq:rob}
\end{equation}
Here, $\eta$ is the momentum ratio of the two stellar winds in the
system and D is the binary separation. For our standard model,
$\eta=0.1$ and $r_{\rm OB}/D= 0.24$.  Assuming a strong shock, and that the
emission is concentrated near the shock apex within an off-axis  distance of
$\pi r_{\rm OB}/2$\footnote{This is true in the wider systems, but in close systems
strong absorption along the line of sight to the shock apex can create
flux maxima to either side of the shock apex, as shown in
Fig.~\ref{fig:dsep_com}. However, in such cases a substantial
proportion of the mass flux (relativistic electron flux) through these
regions will have passed through (originated in) the shock at some
non-negligible off-axis distance. Therefore, the following argument
should be good to first order, even in close systems.}, the characteristic
flow timescale for gas is $t_{\rm flow}=2 \pi r_{OB}/v$, where
the post-shock velocity is $v/4$, with $v$ being the relevant
pre-shock velocity of either the WR or OB stellar wind.

For relativistic particles with $v/c \approx 1$, the rate of energy loss
through IC cooling is \citep{Rybicki:1979}

\begin{equation}
\label{eq:ic_loss}
\left.\frac{dE}{dt}\right|_{\rm IC} = \frac{\sigma_{T} \gamma^{2}}{3 \pi} 
\frac{L_{*}}{r^{2}},
\end{equation}

\noindent where $L_{*}$ is the stellar luminosity and $r$ is the
distance of the shock from the star.  Since the relative momentum
fluxes of the WR and OB stellar winds normally cause the collision
zone to be closer to the OB star, which is generally also the more
luminous star in the binary, $L_{*}$ and $r$ should be evaluated for
the OB star.  From Eqn.~\ref{eq:ic_loss}, the timescale for 50\%
energy loss through IC scattering is

\begin{equation}
\label{eq:t_ic}
t_{\rm IC} = \frac{3 \pi m_{e} c^{2} r_{\rm OB}^{2}}{\sigma_{\rm T} L_{*} \gamma},
\end{equation}

\noindent where we have assumed that to first order the average
distance of the relativistic electrons from the source of UV photons
during this time is $r_{\rm OB}$. If the distance from OB star (the
dominant source of UV photons) to the relativistic electrons is such
that an electron of energy $\gamma m_{e} c^{2}$ loses 50\% of its energy
through IC cooling during the flow time, then $t_{\rm flow} = t_{\rm
ic}$ when

\begin{equation}
\label{eq:D_ic}
r_{\rm OB} = \frac{2 \sigma_{\rm T} L_{*} \gamma_{\rm flow}}{3 m_{e} c^{2} v },
\end{equation}

\noindent or alternatively, when

\begin{equation}
\label{eq:gamma_ic}
\gamma_{\rm flow} = \frac{3 m_{e} c^{2} v r_{\rm OB}}{2 \sigma_{\rm T} L_{*}}.
\end{equation}

\noindent If we assume further that the luminosity of the OB star is
$L_{*} = 10^{5} \Lsol$, then from Eqn.~\ref{eq:gamma_ic} we find
$\gamma \approx 500$ for our standard model, which from
Eqn.~\ref{eq:nu_c} gives a characteristic frequency $\nu_{c} \approx
6$~GHz for $B = 6$~mG.  Hence, for our standard model, the above order
of magnitude estimate suggests that IC cooling starts to have an
effect on the relativistic electrons with $\gamma > \gamma_{\rm
flow}$. In practice, this is likely to represent a lower limit.
Firstly, it has been derived for electrons accelerated at the symmetry
axis and which then flow to a distance of $\pi r_{\rm OB}/2$ from the
symmetry axis.  Electrons accelerated somewhat off-axis will, of
course, take less time than $t_{\rm flow}$ to reach this off-axis
distance. Moreover, the average distance to the OB star will be
greater than $r_{\rm OB}$. Therefore, we expect that IC cooling has
only a significant effect for $\gamma$ greater than a few times
$\gamma_{\rm flow}$. For our standard model, electrons with
$\gamma\sim 3\gamma_{\rm flow}$ result in emission with a
characteristic frequency of $\sim50$~GHz.

This estimate suggests that IC cooling is an important consideration
in {\em all} systems, regardless of binary separation, for
sufficiently high frequencies (or $\gamma$).  However, it is clearly most
significant in shorter period systems, since $\gamma_{\rm flow}
\propto D$ and $\nu_{c} \propto \gamma^2 B \propto D$ (assuming $B
\propto 1/D$). Models with IC cooling included explicitly will be
investigated in Pittard et al. (in preparation).

\subsection{Effect of inclination angle}
\label{sec:inc}

\begin{figure}
\vspace{6.15cm}
\includegraphics{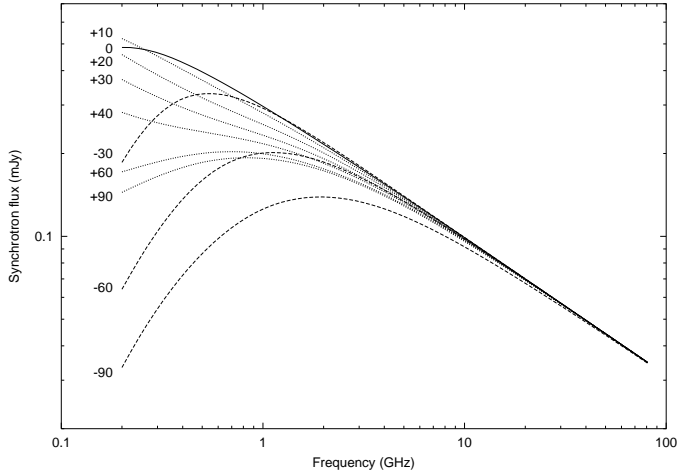}
\caption[]{The effect of free-free opacity as a function of
inclination angle on the synchrotron spectra of the standard
model with $\zeta=10^{-4}$. The inclination angle of the axis of
symmetry to the plane of the sky for the various spectra is
shown. Dashed lines indicate $i< 0^\circ$ (\ie WR star in front), the
solid line $i = 0^\circ$ (\ie quadrature), and dotted lines
$i > 0^\circ$ (\ie OB star in front).}
\label{fig:inc_spec}
\end{figure}

We now examine the effect on the observable synchrotron spectrum of
changing the inclination of {the axis of symmetry (the line of centres)
to the line of sight using our standard model.} This is shown in
Fig.~\ref{fig:inc_spec}, where only the effect of the free-free
opacity of the circum-binary stellar wind envelope is being
considered. Since our model is axis-symmetric, changing only the
inclination covers {\em any} orientation of the system relative to the
observer. The largest effects are seen at the lowest frequencies,
where the free-free opacity is highest for a given path length through
the circum-binary envelope ($\tau_{\rm ff}\propto \nu^{-2.1}$).  At an
inclination angle $i = 0^\circ$, the lines-of-sight are perpendicular
to the axis of symmetry of the model; at $i = -90^\circ$ they pass
first through the WR-star wind; and at $i = +90^\circ$ they pass first
through the OB-star wind. The asymptotic half-opening angle of the
WR-shock cone is $72^{\circ}$, so when $i > 18^{\circ}$, lines of
sight into the system can impinge the shocked region without first
passing through unshocked circum-binary gas.

\begin{figure}
\vspace{15.6cm}
\includegraphics{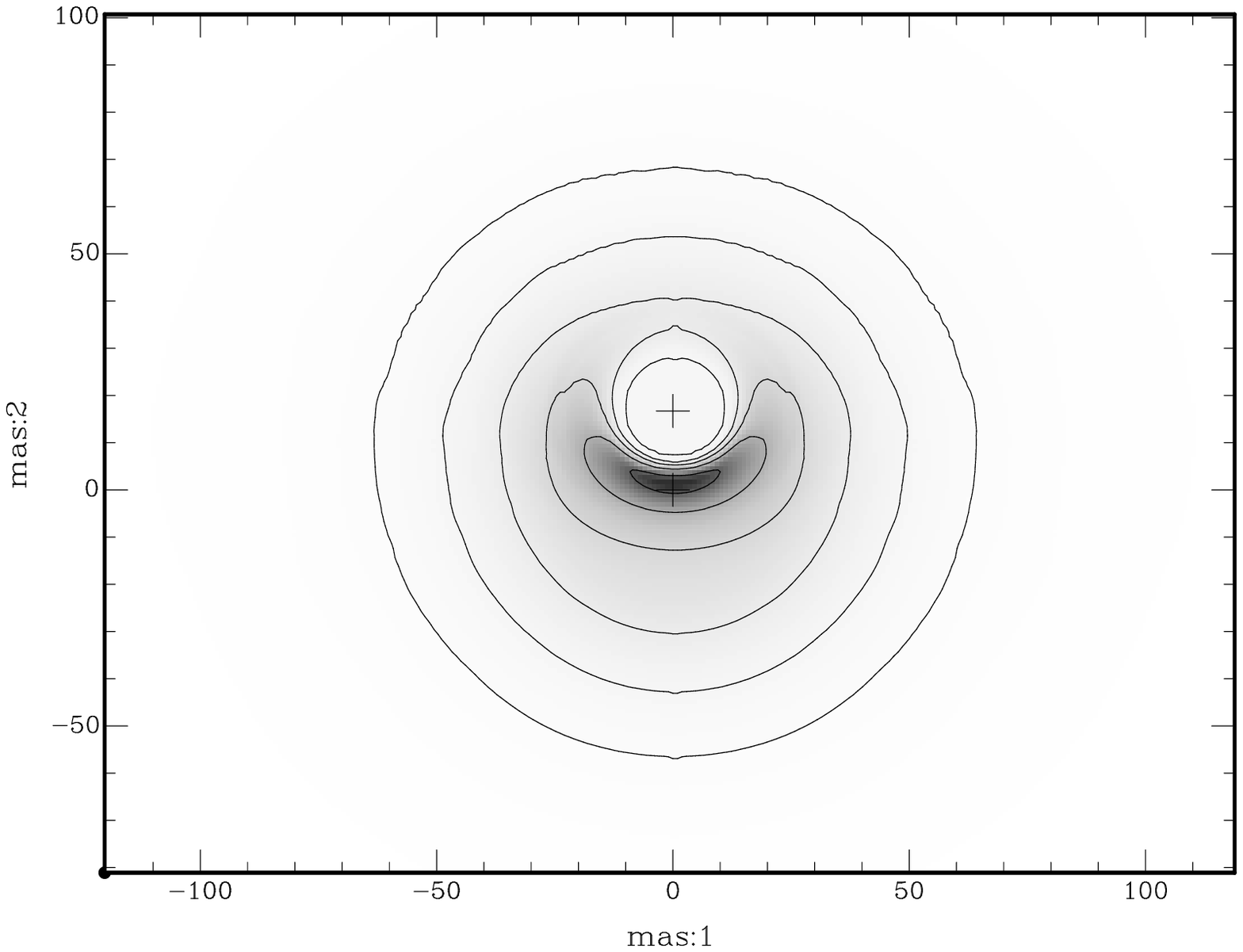}
\includegraphics{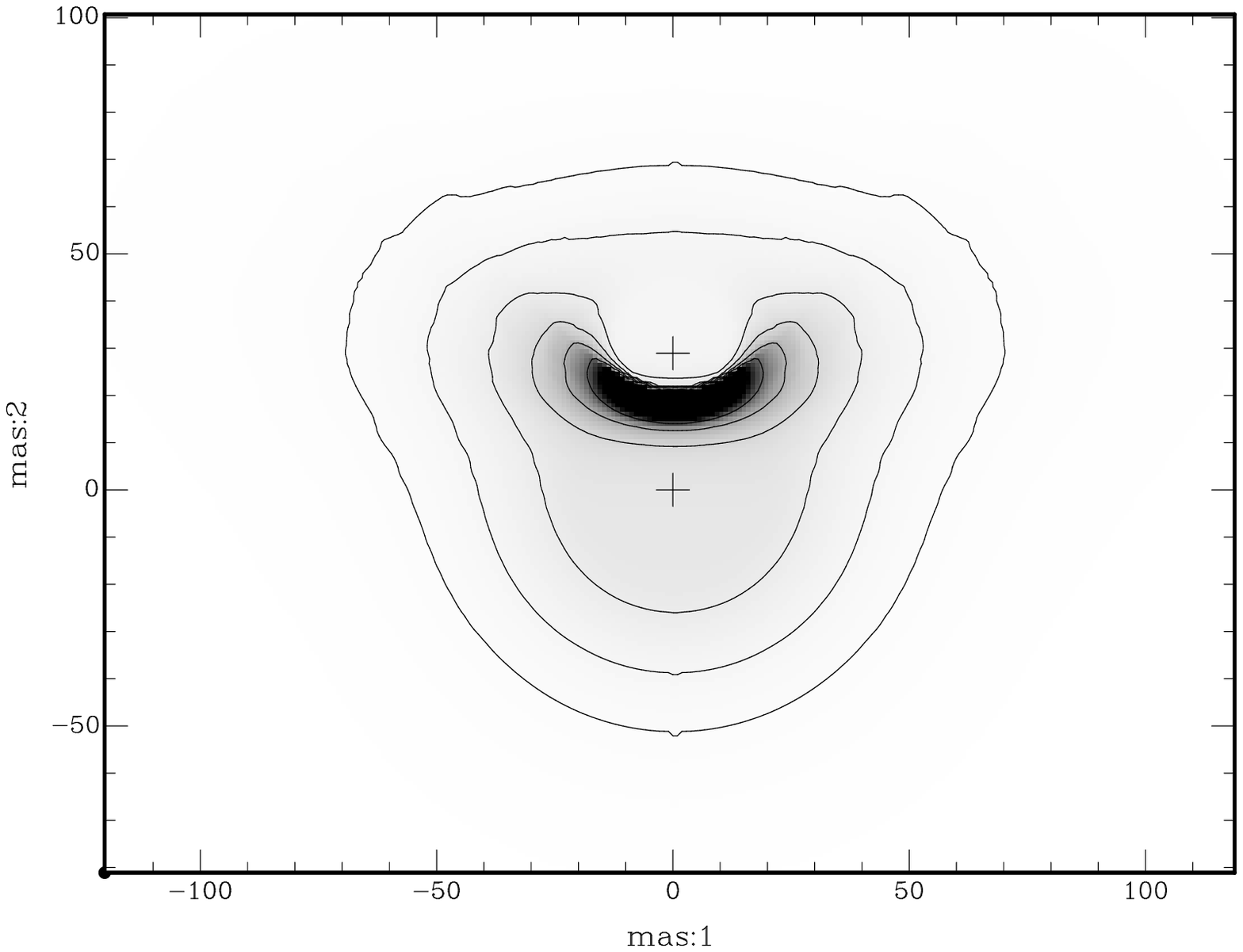}
\includegraphics{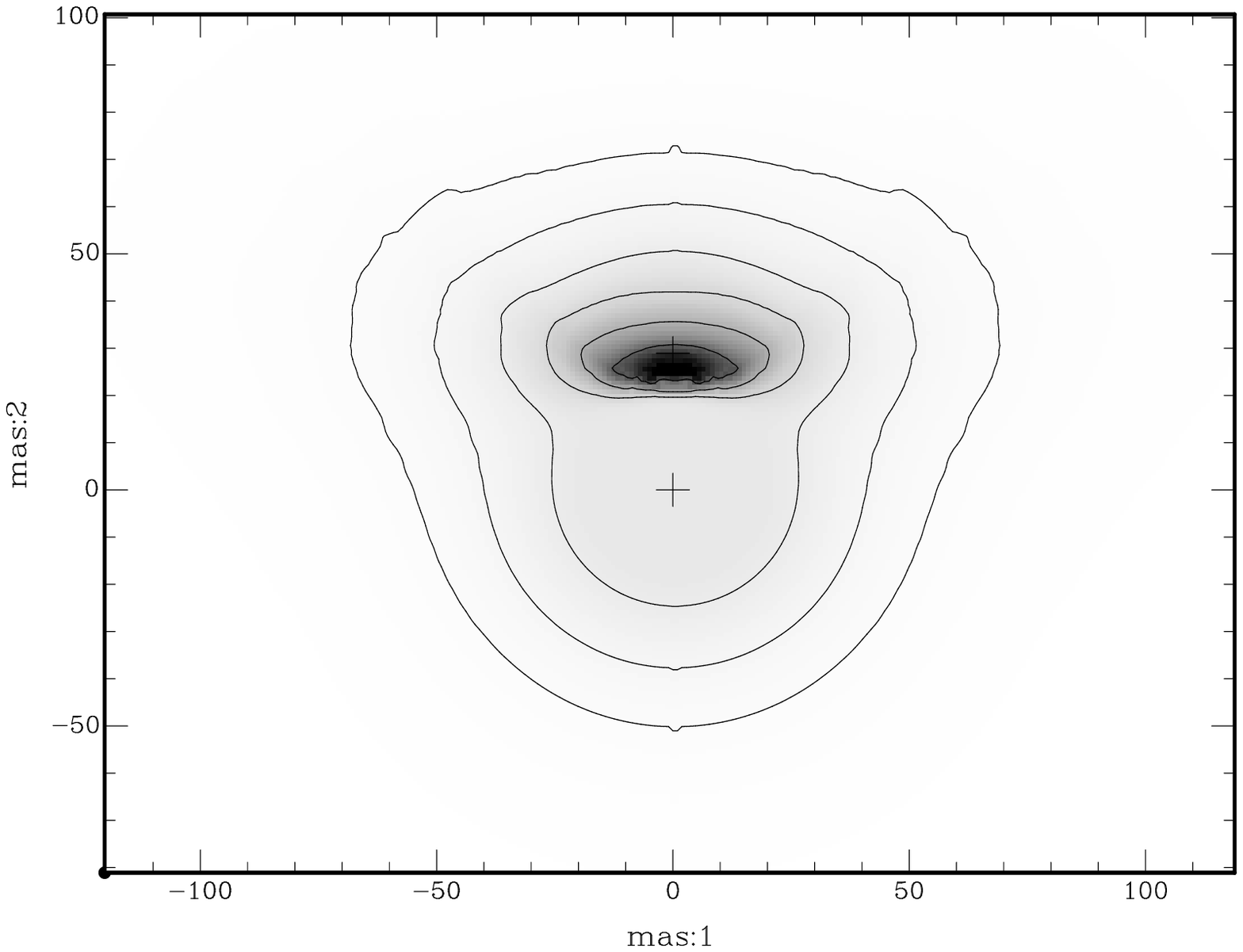}
\caption[]{The 1.6-GHz intensity distribution for a model with $D = 5
 \times 10^{14}$~cm at $+60^\circ$ (top), $+30^\circ$ (middle), and
 $-30^\circ$ (bottom) inclination. At positive angles the OB star is
 closer to the observer than the WR star.  The crosses denote the
 position of the stars, with the WR star at (0,0). Each image has the
 same intensity greyscale and contour intervals (also used in the
 middle panel of Fig.~\ref{fig:dsep_com}, showing the $0^\circ$ case).
 The optically thick part of the OB-star wind is the obvious ``hole''
 (top panel), below which the synchrotron emission is seen, surrounded
 by the free-free emission, largely from the WR wind.  As inclination
 decreases, the hole becomes less pronounced and the synchrotron
 intensity increases as the line-of-sight opacity through the OB wind
 decreases. When the WR star is closer to the observer ($i<0$), the
 optically thick part of the WR-star wind obscures the lower part of
 synchrotron emission region.}
\label{fig:image_inc}
\end{figure}

The effects of changing free-free opacity with inclination are
clear. The size of the optically thick region of the circumbinary
nebula goes as $\nu^{-0.7}$ for an $r^{-2}$ radial density gradient,
and so as frequency increases the opacity is reduced along any given
line-of-sight.  In our standard model, it is evident that the
cirumbinary nebula is optically thin above a few~GHz.
The largest free-free attenuation occurs when the line of sight to the
wind-wind collision region passes closest to the WR star \ie at
$i=-90^\circ$. As the inclination angle increases, the line of sight
passes through less dense regions of the WR star and the opacity is
reduced, resulting in higher observed synchrotron flux.  For $i =
0^\circ$ the lines of sight to the wind-wind collision region pass
largely through the outer regions of the WR star envelope and the low
free-free opacity of the wind-wind collision region. For $i >
0^{\circ}$, the region of the wind-wind collision zone occulted by the
optically thick surface of the OB-star wind moves gradually toward the
shock apex as $i$ increases (\cf Fig.~\ref{fig:image_inc}). This, and
the fact that the lines of sight to the shock apex pass through the
dense inner regions of the OB-star wind, is the cause of the
increasing absorption at low frequencies with increasing $i$. The flux
when $i = +90^{\circ}$ is always greater than the flux when $i =
-90^{\circ}$ since the optical depth unity surface of the WR wind is
always larger than that of the OB-star wind owing to the higher
density of the WR wind.

The effect of inclination on the resulting intensity distribution is
demonstrated in Fig.~\ref{fig:image_inc}, where the line-of-sight
passes down the shock cone, and the OB-star wind absorbs flux from its
far side. Absorption by the optically thick region of the OB-star wind
is distinctly apparent. Clearly the inclination angle of a given
source can have a large effect on the resulting flux and the spatial
intensity distribution.

\begin{figure}
\vspace{6.15cm}
\includegraphics{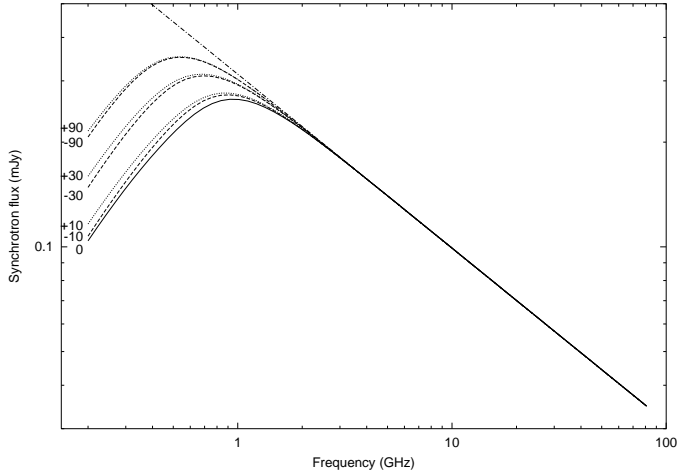}
\caption[]{The effect of SSA on the spectrum of the emission emerging
from the shocked region as a function of inclination angle for
the standard model with $\zeta=10^{-4}$. The inclination angle of the
axis of symmetry to the plane of the sky for the various spectra is
shown. Dashed lines indicate $i< 0^\circ$ (\ie WR star in front), the
solid line $i = 0^\circ$ (\ie quadrature), and dotted lines $i >
0^\circ$ (\ie OB star in front). The dot-dashed line is the intrinsic
synchrotron emission from the shocked gas, which is independent of
inclination, plotted for reference.}
\label{fig:inc_ssa_int}
\end{figure}

\begin{figure}
\vspace{6.15cm}
\includegraphics{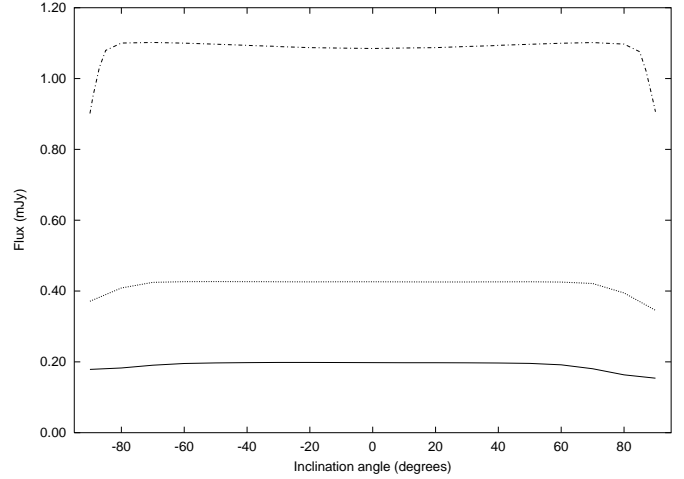}
\caption[]{The change in thermal flux as a function of inclination angle
using the standard model with $\zeta=10^{-4}$. Shown are curves for 
1.6 (solid), 5 (dotted) and  22 GHz (dot-dashed).}
\label{fig:inc_ffnt}
\end{figure}

The Razin effect only affects the generation of synchrotron photons
and is independent of inclination.  On the other hand, SSA is
dependent on the path length through the shocked gas, and inclination
angle influences its effect on the low frequency synchrotron
spectrum. This is shown in Fig.~\ref{fig:inc_ssa_int}, where the
intrinsic synchrotron spectrum is shown. Examining the intrinsic
spectrum eliminates the effect of the large changes in free-free
opacity with inclination angle, as seen in
Fig.~\ref{fig:inc_spec}. The dependence of SSA on the inclination
angle is clearly a function of the path length through the shocked
envelope, particularly the highest density (shock apex region) part of
the shocked gas. Thus at high angles, this path length is small and
the impact of SSA on the spectrum is lowest, whereas at quadrature the
lines-of-sight through the shock apex region are at maximum length and
the impact of SSA on the spectrum is highest. However, the effect is
not as dramatic a function of inclination angle as the effect of
varying free-free opacity with inclination angle.

When the Razin effect, SSA, and free-free opacity are considered
together in our models it is important to bear in mind that both the
Razin and SSA are functions of $\zeta$ (see
Sec.~\ref{sec:raz_ssa}). As an example, when $\zeta=10^{-4}$, the
Razin effect is so dominant that the spectrum is influenced little by
changes in the inclination angle. The changes that do occur are due
mostly to the changing free-free opacity.

In Fig.~\ref{fig:inc_ffnt} we show how the thermal component varies
with inclination angle in our standard model.  The observed free-free
flux is essentially constant throughout the range of inclinations,
except when $i$ approaches $-90^\circ$ and $+90^\circ$. At these
extreme inclinations, the optically thick parts of, respectively, the
WR and OB star winds absorb the free-free emission arising from the
bulk of the stellar wind of the other star and from the wind-wind
collision region (a relatively small contribution to the total thermal
emission \citep{Stevens:1995}), producing a decrease in thermal flux.
The free-free flux at $i = -90^\circ$ is slightly higher than at $i =
+90^\circ$, consistent with the higher density of the WR star wind.

\section{Modelling the radio emission from \object{WR\,147}}
\label{sec:147}
Having explored how the radio flux varies with separation and
inclination in the previous section, we now turn our attention to the
modelling of a specific system.  \object{WR\,147} is notable for being among
the brightest WR stars at radio frequencies, and for being one of two
systems in which the thermal and synchrotron emission are observed to
arise from two spatially resolved regions \eg~\citet[and references
therein]{Williams:1997}. Furthermore, it is one of a handful of WR+OB
binary systems where the two stars are spatially resolved
\citep{Williams:1997, Niemela:1998}. These observations suggest a very
wide system with a projected separation $D\cos~i=0.635\pm0.020$\arcsec.
At the estimated distance of $\sim0.65$~kpc \citep{Churchwell:1992,
Morris:2000} this corresponds to a separation
$D\sim415/\cos~i$~AU. This relationship between $D$ and $i$ represents
an important constraint for any models of the system. The inclination
angle is unknown so we investigate models for several different values
which requires different values of the physical separation, $D$, to
maintain the observed angular separation. This leads to different
sets of physical parameters, including the normalisation constant
$\zeta$, to fit the observed spectrum. 

\subsection{Modelling the radio spectrum of \object{WR\,147}}
\begin{figure}
\vspace{6.15cm}
\includegraphics{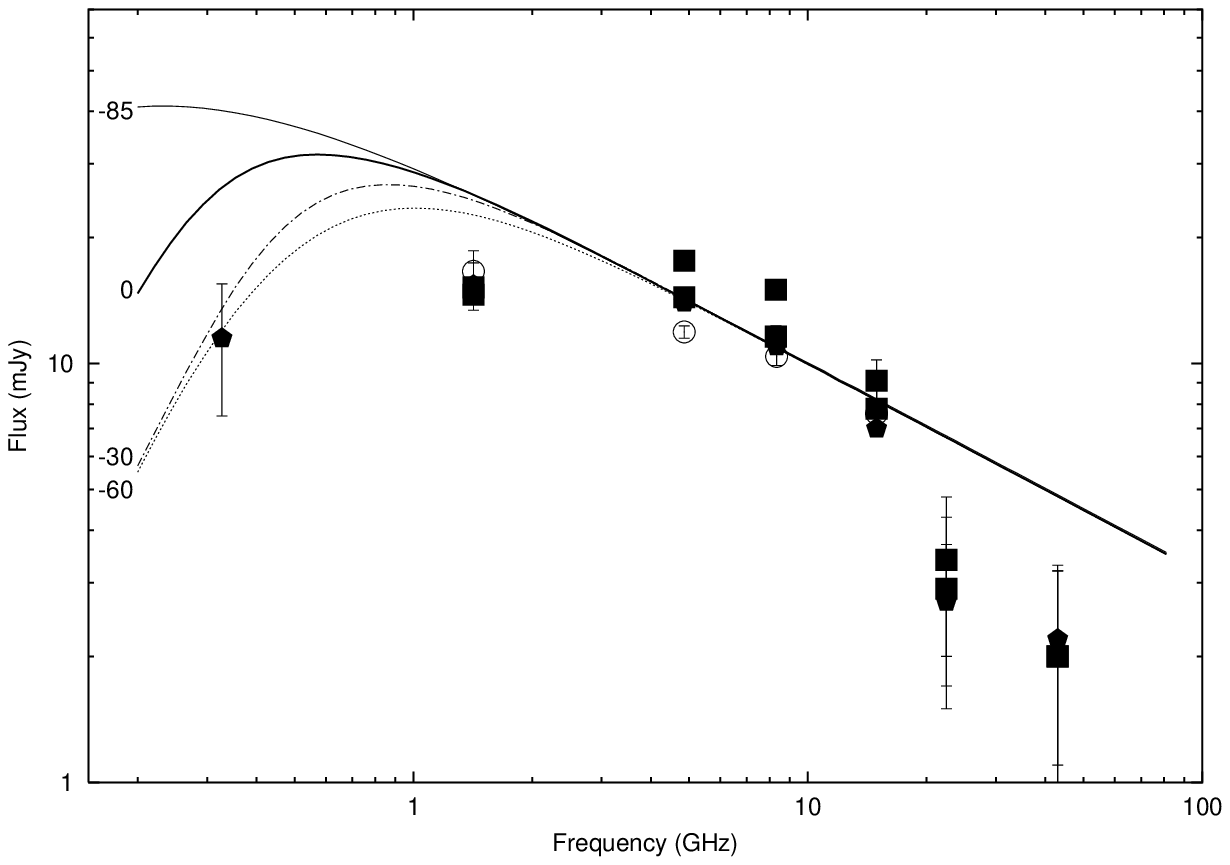}
\caption[]{The synchrotron spectrum of \object{WR\,147} as deduced by several
authors - solid squares (two separate estimates from the same
observational data by~\citet{Skinner:1999}), solid pentagons
\citep{SetiaGunawan:2001b}, open circles (fluxes from
\citet{Churchwell:1992} and~\citet{Contreras:1999}). The flux at both
22 and 43 GHz is highly uncertain, with the latter estimated from a
long extrapolation of a power-law fit to the lower frequency
``thermal'' data points to be $\approx2$~mJy~\citep{Skinner:1999,
SetiaGunawan:2001b}. Both high frequency points suggest a high
frequency turn down.  The lines are theoretical spectra from models
where only free-free absorption in the circum-binary stellar wind
envelope is attenuating the synchrotron emission. Models for various
inclination angles are shown: $0$, $-30$, $-60$ and $-85^\circ$
(corresponding to $\zeta=6.92, 7.23, 8.08$, and $13.44\times10^{-3}$
respectively). The $0^\circ$ model is denoted by the solid line.}
\label{fig:1st_fits}
\end{figure}

The radio spectrum of \object{WR\,147} is perhaps the best observed of
all massive binary systems, with radiometry extending from 353~MHz to
42.9~GHz.  Reviewing the literature, it is immediately apparent that
differences in excess of 50\% exist in the derived synchrotron flux at
some frequencies (cf. Fig.~\ref{fig:1st_fits}), giving rise to
uncertainty in the position of the low frequency turnover.
\cite{Skinner:1999} estimate that at 5~GHz the synchrotron flux
($S_{\rm 5GHz}$) is greater than the synchrotron flux at 1.4~GHz,
whereas \cite{SetiaGunawan:2001b} estimate that $S_{\rm 1.4GHz} >
S_{\rm 5GHz}$. These differences may be attributed to a number of
factors that include analysis techniques, resolution effects, the
assumed thermal contribution, and possibly temporal variations. {
Though \cite{Churchwell:1992} report variations in radio flux
approaching 50\%, subsequent reanalysis of the same data by
\cite{SetiaGunawan:2001b} suggests that any variations are much
smaller ($\sim15$\%) and are long-term, with little power on time
scales of days.  \cite{Skinner:1999} completed their observations of
\object{WR\,147} within a couple of days, and their repeated
observation at 1.4~GHz showed no variation. This suggests that, at the
very least, their observed spectrum is free of temporal variations.}
In addition, it is clear there are difficulties in determining the
component fluxes when the sources are spatially resolved \cf Table~1
in~\citet{Skinner:1999} and Table~3 in~\cite{Churchwell:1992}.

These differences form the fundamental reason for conflicting
conclusions about the nature of the underlying electron energy
spectrum in the current literature. While both \cite{Skinner:1999} and
\cite{SetiaGunawan:2001b} note that a free-free absorbed synchrotron
power-law model is a poor fit to the data, \cite{SetiaGunawan:2001b}
argue that the deficit in the observed flux above 15~GHz is due to a
high energy limit to the electron acceleration. In contrast,
\cite{Skinner:1999} argue that the spectrum is best fit by a {\em
mono-energetic} relativistic electron distribution. At high
frequencies, the spectrum from such a distribution falls off as
$P(\nu) \propto \nu^{1/2} e^{-\nu}$. This fits the 15 and 22 GHz data
very well, but the flux at these frequencies are most uncertain.
Furthermore, the fit of a mono-energetic electron distribution spectrum
to the high frequency fluxes depends also on the position of the low
frequency turnover which, as already noted, is uncertain. 
How such an electron distribution results from a shock acceleration
is unclear.

Using our model, we attempt to model the spectrum of \object{WR\,147}
assuming a power-law electron energy distribution. For the stellar
wind parameters we adopt $\Mdot_{\rm WN8} = 2 \times 10^{-5}
\Msolpyr$, $v_{\rm \infty WN8} = 950 \kmps$, $v_{\rm \infty OB} = 1000
\kmps$ and a wind momentum ratio $\eta = 0.02$
\citep{Pittard:2002b}. The last is toward the higher end of reasonable
estimates, and since it gives $\Mdot_{\rm OB} = 3.8 \times 10^{-7}
\Msolpyr$ favours a companion of spectral type around late-O or
early-B supergiant. Such parameters are consistent with the estimated
B0.5 by~\citet{Williams:1997} based on an uncertain luminosity
estimate, and the more recent O5-7I deduced from HST STIS spectroscopy
\citep{Lepine:2001}. The composition of the WN8 stellar wind is taken
from~\citet{Morris:2000}, giving $X=0.09$, $Y=0.89$, and
$Z=0.016$. The wind temperature for both stellar winds was assumed to
be {10~kK, and the dominant ionization states were H$^+$, He$^+$
and CNO$^{2+}$. The thermal flux is only weakly dependent on the
assumed wind temperature (through the Gaunt factor), if the ionization
state is fixed.}

\begin{figure}
\vspace{6.15cm} \includegraphics{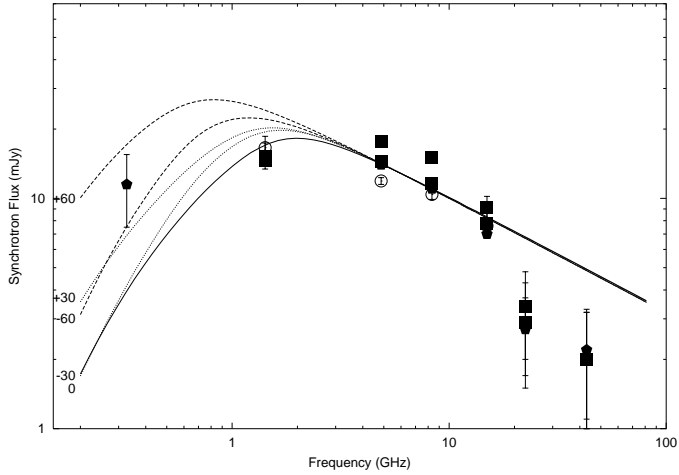}
\caption[]{Model synchrotron spectra when SSA, the Razin effect and
free-free absorption in the circum-binary stellar wind envelope are
included, for various inclination angles: $0$ (solid), $\pm30$
(dotted), and $\pm60$ (dashed), with corresponding values of $\zeta =
(7.03, 7.32, 8.12)\times 10^{-3}$. The data points are those shown in
Fig.~\ref{fig:1st_fits}.}
\label{fig:raz_ssa}
\end{figure}

Values of $\zeta$ are estimated for each of our models by matching the
model to the mean of the ``observed'' synchrotron 4.86~GHz fluxes, 
which we take to be $14.1\pm0.3$~mJy. We then determine the spectrum 
between 0.2 and
80~GHz assuming this constant value of $\zeta$. As the inclination of
the model increases, the binary separation $D$ has to increase in
order to satisfy the observed constraint on $D\cos i$, forcing the
need to increase $\zeta$ to maintain the 5 to 15~GHz flux level.

{To fit the thermal flux with our chosen value of $\Mdot$ we
require the winds to be clumped. This has some precedent:
\citet{Morris:2000} note that the observed thermal flux implies that
$\Mdot \sim 7.5 \times 10^{-5}\;\Msolpyr$ if the wind is assumed
non-clumpy, but find that models with a homogeneous wind cannot match
the IR helium-line profiles observed with {\it ISO}. Their derived
volume-filling factor of $f \sim 0.1$ implies an actual mass-loss rate
$\Mdot \sim 2.4 \times 10^{-5}\;\Msolpyr$. We have therefore
multiplied $\varepsilon_{\nu}^{\rm ff}$ and $\alpha_{\nu}^{\rm ff}$
from Eqns.~1 and~2 by a factor $1/f$ to simulate clumping in the cool
stellar winds. We assume that both the WR and O-star winds are clumped
by the same factor. Also, it is assumed that the clumps are destroyed
as they pass through the shocks into the wind collision zone, and
therefore do not increase the free-free emission from this region.  We
adopt $f = 0.134$\footnote{Interestingly, this implies that the
mass-loss rate of the equivalent homogeneous wind is $\Mdot =
\Mdot_{\rm actual}/\sqrt{f} = 5.5 \times 10^{-5} \Msolpyr$.  While
this is below the observationally deduced value using a single star
model, it is consistent with the 13-38\% enhancement in $S_{\nu}$
expected from a binary with a wind momentum ratio $\eta = 0.1$
\citep{Stevens:1995}.}  to ensure that the thermal flux at 22GHz
remains less than the total flux. }

{Fig.~\ref{fig:1st_fits} demonstrates that the low-frequency
turnover could potentially be accounted for in our models by free-free
opacity of the circum-binary stellar wind at intermediate inclination
angles, with the line of sight into the system through the WR
wind\footnote{\citet{Dougherty:2003} erroneously report a fit to the
low-frequency spectrum of \object{WR\,147} using only free-free
absorption. Their fit was attained using SSA, in addition to free-free
absorption.}, but not at $0^\circ$ or inclinations angles for which
the lines-of-sight to the shock apex pass through the densest parts of
the WR wind ie. $-85^\circ$. The models do not fit the 1.4~GHz fluxes
particularly well.} \citet{SetiaGunawan:2001b} arrived at a different
conclusion: that free-free absorption alone was sufficient to account
for the turnover, based on results from a synchrotron point-source
model. We attribute our different conclusion to the extended nature of
the region of shocked material in our model; although the
line-of-sight to the shock apex may be optically thick, lines-of-sight
to the outer regions of the wind-wind-collision region may remain
optically thin. Maximum free-free absorption of the low frequency
synchrotron emission occurs around inclination angles of
$-60^{\circ}$. For still lower inclination angles, the observed flux
is higher despite increased attenuation to the shock apex. This is
because increased $\zeta$ leads to synchrotron emission further from
the shock apex. The only manner in which free-free opacity {\em alone}
could produce a sufficiently strong low-frequency turnover would be if
the size of the synchrotron emission region were reduced, such that
the greater attenuation near the shock apex became more significant to
the observable emission. Such a reduction in the size of the emission
region could be mimicked in an ad-hoc fashion in our model by evolving
$\zeta$ away from the shock apex. Certainly, as the shocks become more
oblique moving away from the shock apex, the particle injection and
acceleration efficiency is lowered
\citep{Ellison:1995,Ellison:1996}. The energy spectrum is also
evolving since IC cooling is ongoing as the electrons advect away,
though this predominantly effects the higher energy electrons, and
coulombic cooling may be more important for the lower energy
electrons.

\begin{figure}
\vspace{6.15cm} \includegraphics{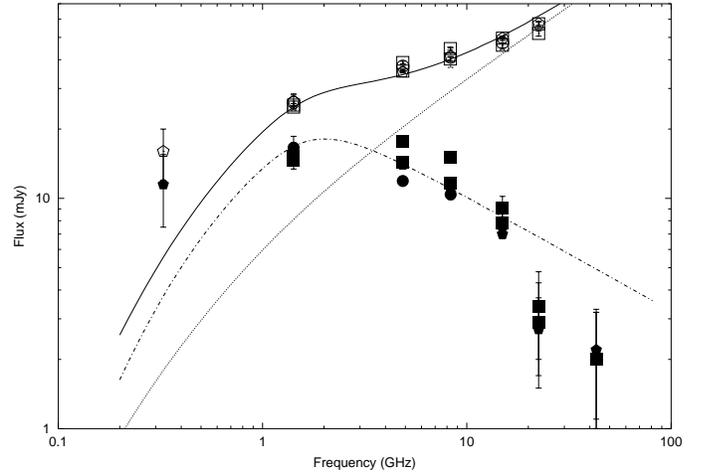}
\caption[]{The spectrum of \object{WR\,147}. The synchrotron data are those
shown in figs.~\ref{fig:1st_fits} and~\ref{fig:raz_ssa} and
represented by solid data points. The total flux as deduced by the
same authors are represented by the hollow data points. The lines are
the total (solid), thermal (dotted) and synchrotron (dot-dashed)
spectra of the $i=0^\circ$ models with $\zeta=7.03\times10^{-3}$,
where SSA, Razin and free-free absorption are included in the
radiative transfer calculations. {The volume-filling factor
$f=0.134$. Models at $+30^\circ$ and $-30^\circ$ are also consistent
with the data.}}
\label{fig:thermal_sync_data}
\end{figure}

\begin{table}
\label{tab:parms}
\caption[]{Values of density, equipartition field, and peak temperature
near the shock apex for the models shown in Fig.~\ref{fig:raz_ssa}.}
\begin{tabular}{llll}
\hline
Inclination  & $0^\circ$ & $30^\circ$ & $60^\circ$ \\
\hline
$D \times10^{-15}$ cm &6.21&7.17&12.4\\
$n_e(\rm WR) \times 10^{-4}$ cm$^{-3}$& 1.8 & 1.35 & 0.45 \\
$n_e(\rm OB) \times 10^{-4}$ cm$^{-3}$& 3.3 & 2.5  & 0.83 \\
$B^1$ mG & 4.1 & 3.6 & 2.2\\
$\zeta \times 10^3$ & 7.03 & 7.32 & 8.12 \\ 
\hline
T$_{\rm max}$ (WR) K& \multicolumn{3}{c}{$2.4\times10^7$}\\
T$_{\rm max}$ (OB) K&\multicolumn{3}{c}{$1.4\times10^7$}\\
\hline
\end{tabular}

$^1$ The B-field at the shock apex is approximately the same
in both the shocked WR and OB regions.
\end{table}

Models that include the influence of both SSA and the Razin effect in
addition to free-free absorption are more successful in fitting the
low-frequency data. In Sec.~\ref{sec:param_study}, both these
effects were shown to be significant, depending on the value of
$\zeta$, and it seems unrealistic to exclude such fundamental
processes from our models. The resulting spectra are shown in
Fig.~\ref{fig:raz_ssa}. As before, values of $\zeta$ have been chosen
for the different models based on matching the 5-GHz data point, and
are $\sim1$\%, similar to the values used in models of SNR
(cf.~\citet{Mioduszewski:2001}) and of an order of magnitude expected
for shock-accelerated electrons (see Sec.~\ref{sec:synch}). As in
the case of free-free opacity alone, as the separation increases,
$\zeta$ increases and the amount of emission arising far from the
shock apex increases. Thus the low-frequency turnover from optically
thin to optically thick emission moves to lower frequencies with
increasing separation. The difference between the spectra for positive
and negative inclinations of the same absolute value is due to the
different free-free opacity through the OB-star wind as opposed to the
WR-star wind. The values of density, equipartition field, and peak
temperature near the shock apex for the models shown in
Fig.~\ref{fig:raz_ssa} are given in table~1.

In Fig.~\ref{fig:thermal_sync_data} the resulting total, synchrotron
and thermal spectra from one of our models ($i=0$, and including SSA,
the Razin effect and free-free absorption) is shown against the
observed spectra. Though this is not the best fit to the data in a
formal sense, our model fits the data reasonably well given the
approximations and assumptions which it contains. 
In the model shown in Fig.~\ref{fig:thermal_sync_data}, the total flux
in the 5 to 8 GHz range is underestimated, and the synchrotron
emission does not turn down somewhere around 10 to 20~GHz to account
for the observed high frequency data.  We could account for the
shortfall in the 5-8~GHz total flux by increasing the thermal emission
by a few mJy. This could be achieved by simply increasing
$\Mdot/v_\infty$ in the stellar winds. However, this would require
that the synchrotron emission at frequencies higher than $\sim10$~GHz
be lower than in the current models, which may well be the case if IC
cooling of the highest energy electrons is taken into account, as
explained below.

Fig.~\ref{fig:thermal_sync_data} lends considerable support to the
suggestion that WR stars with continuum spectra of spectral index
less than $+0.6$ are synchrotron emitters \eg~\citet{Dougherty:2000b}. The
total flux spectrum as shown in Fig.~\ref{fig:thermal_sync_data} could
be fit with a power-law spectrum of spectral index of $\sim +0.3$, 
which  could be interpreted as arising from a partially optically thick
stellar wind. However, it is well-established that the stellar winds
of WR stars have radio continuum spectra with spectral indices
$+0.6\rightarrow +0.8$ \eg~\citet{Williams:1990b, Leitherer:1991}, and
spectral indices less than $\sim+0.6$ are not expected from stellar 
winds alone.  This suggests that spectral indices less than $+0.6$ are
a strong indicator of the presence of a synchrotron component. This is
especially useful in those cases \eg \object{WR\,86}, where the synchrotron and
thermal emission regions are not resolved as separate components, yet
a continuum spectrum with a low spectral index can be observed. 

We are unwilling to draw much of a conclusion related to inclination
angle from these models of the spectrum of \object{WR\,147}. For models with
wider separations \ie higher inclinations, the 1.4 GHz data is
overestimated substantially, which hints at an inclination angle
somewhere between $0^\circ$ and $30-40^\circ$. The high end of this
estimate is consistent with the poorly constrained inclination
estimates from~\cite{Williams:1997}
and~\cite{Contreras:1999}. However, any conclusion related to
inclination angle drawn from fitting the spectrum has to be tempered
due to the heavy reliance on the accuracy of the 1.4 GHz data points,
and the poor precision of the 352 MHz data point.

Another striking feature of
Figs.~\ref{fig:1st_fits},~\ref{fig:raz_ssa}
and~\ref{fig:thermal_sync_data} is our inability to generate a
high-frequency turnover somewhere around 10 to 20 GHz, as suggested by
the data points. This is a consequence of using Eqn.~\ref{eq:sync} to
describe the synchrotron power, where it is assumed that $\gamma_{\rm
max}$ is infinite. IC cooling first determines the maximum energy that
can be achieved by the acceleration processes. As the relativistic
electrons are advected away from the acceleration site by the flow,
the highest energy particles suffer crippling losses from IC cooling
as this is no longer balanced by mechanisms supplying rapid energy
gain. Hence $\gamma_{\rm max}$ will have a finite value. For \object{WR\,147},
the rate of energy gain by 1st-order Fermi acceleration is equal to
the rate of energy loss by IC cooling when $\gamma = \gamma_{\rm max}
\sim 10^6$. From Eqn.~\ref{eq:ic_loss}, electrons with $\gamma =
10^{6}$ are cooled to $\gamma_{\rm flow} \sim 400$ within a time
$t_{\rm flow}$ when $r \approx r_{\rm OB}$ and $L_{*} = 10^{5}
\Lsol$. The high frequency turn down seen in the observational data
occurs at $\sim 10-20$~GHz, which for $B \sim 3$~mG suggests $\gamma
\sim1200$, which is within a factor of a few of the value estimated
above. Thus, we conclude that the high frequency turn down is an
expected consequence of IC cooling. {This break in the energy
spectrum of the relativistic electrons will produce a corresponding
break in the spectrum of the IC photons. For relativistic particles,
the final energy ($E_f$) of an IC photon is related to its initial
energy ($E_i$) by $E_f=\gamma^2 E_i$. For an OB-type star, the stellar
photon distribution peak occurs $\sim10$~eV. This implies that the IC
photons resulting from scattering off electrons with $\gamma\sim10^3$
or greater will have energies in excess of 10~MeV - the hard X-ray and
gamma ray regimes.  Such high energy photon production in CWBs has
been examined by other authors \eg~\citet[and references
therein]{Benaglia:2003}.} An additional point is that since $d\gamma
\propto \gamma^2$, the higher energy electrons cool most rapidly. This
causes the electron distribution to ``bunch-up'' near $\gamma_{\rm
flow}$, and may help to explain why a mono-energetic fit works well. A
model with an evolving value of $\gamma$ will be explored by Pittard
\etal (in preparation).

\begin{figure*}[t]
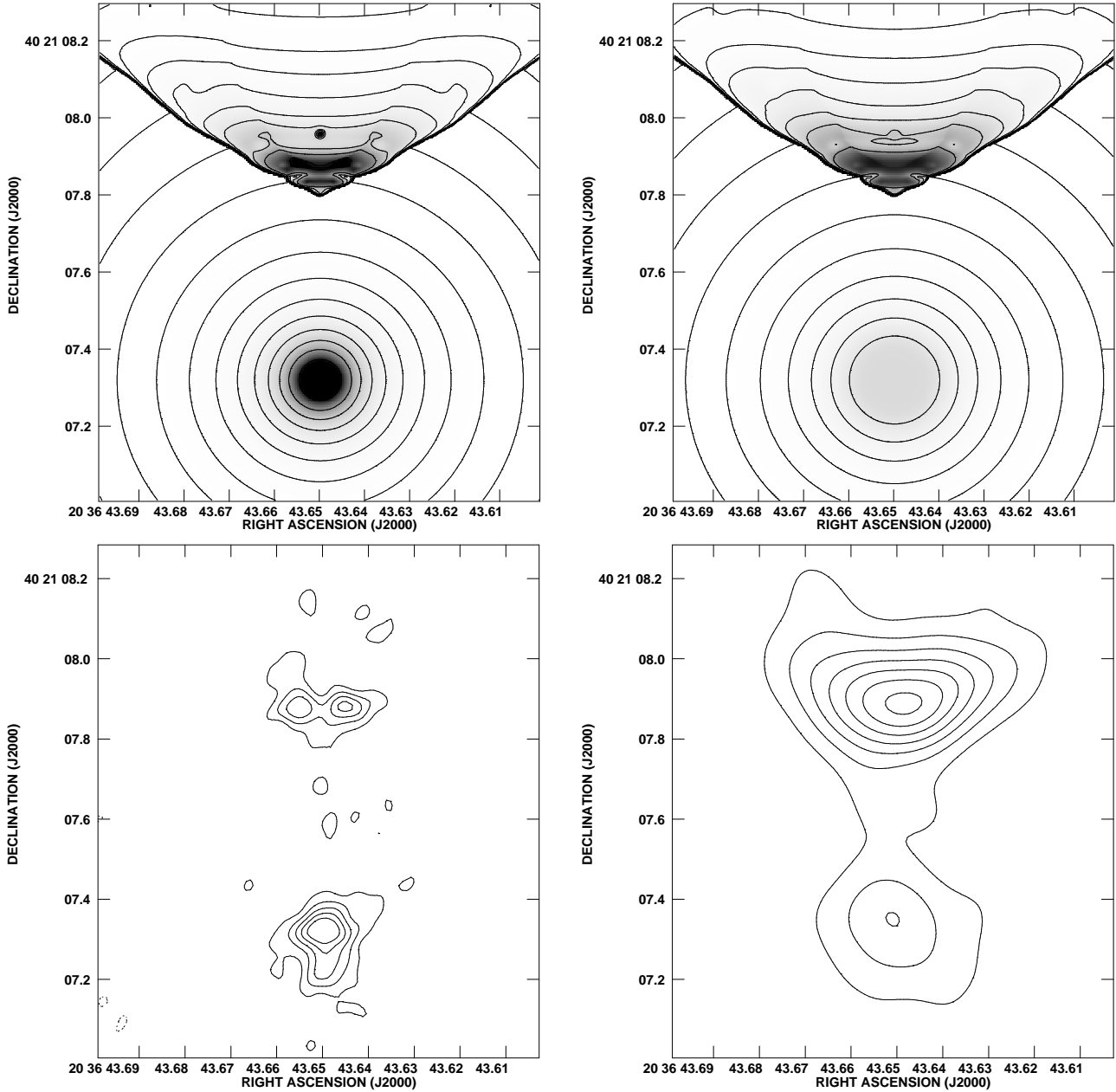

\vspace{16.7cm}
\includegraphics{5ghz_greys.ps}
\includegraphics{1.6ghz_greys.ps}
\includegraphics{5ghz_merlin.ps}
\includegraphics{1.6ghz_merlin.ps}
\caption[]{Intensity distributions at 4.8 (left) and 1.6~GHz (right)
of our \object{WR\,147} model at $0^\circ$ inclination.  The top panels are the
model intensity distributions at each frequency. The greyscale and
contour range are the same in both images. The stellar wind of the
OB-star companion is visible, particularly at 5 GHz, along with the WR
star wind and the wind-wind collision region. The large contour range
was used to show that the stellar wind of both stars extends far
beyond the greyscale range. The lower images are simulated MERLIN
observations generated using the model intensity distributions shown
at the top, and the same $u-v$ distribution and beam sizes 
as~\citet{Williams:1997}. The beams are circular, of diameter 57 and
175~mas at 5 and 1.6 GHz respectively. The contour levels at the two frequencies 
are also the same as those in~\citet{Williams:1997}. 
The similarities to the observations
of~\citet{Williams:1997} is striking, most particularly the emission
from the wind-collision region.}
\label{fig:147_model_images}
\end{figure*}

\subsection{Modelling the MERLIN images of \object{WR\,147}}

The spatial distributions of radio emission at both 5 and 1.6 GHz from
the model in Fig.~\ref{fig:thermal_sync_data} are shown in the top
panels of Fig.~\ref{fig:147_model_images}. At 5~GHz the peak intensity
of thermal emission from the WR stellar wind and the synchrotron
emission from the collision region are similar, and the OB-star
companion is clearly visible. At 1.6~GHz the synchrotron emission is
much brighter than the WR star stellar wind, and the OB-star wind is
barely visible.

To appreciate how these model intensity distributions compare with the
MERLIN observations shown in~\citet{Williams:1997}, we used the AIPS
subroutine UVCON to generate visibilities appropriate for a MERLIN
``observation'' of our models. In addition to the array co-ordinates,
system noise estimates were included in the calculation by including
the performance characteristics of the MERLIN telescopes \eg antenna
efficiencies, system temperatures etc. The resulting visibilities were
then imaged and deconvolved using the same procedure
as in~\citet{Williams:1997}, giving the ``simulated'' observations shown
in the lower panels of Fig.~\ref{fig:147_model_images}. The remarkable
similarity between these images and the observations presented
in~\citet{Williams:1997} is striking. However, here the 5-GHz peak
intensity of the collision region is a little lower than that of the
WR stellar wind, and the WR star is a little too bright at
1.6~GHz. The WR star also appears to be a little more compact than
shown in~\citet{Williams:1997}, but we attribute these minor
differences to our simple spherical, isothermal model of the stellar
winds.  Note that the OB star is not visible in the simulated
observations. This is because the observations are essentially the
intensity distribution (the upper panels) convolved with the
interferometer beam and the emission from the OB star is then
sufficiently dispersed that it is no longer visible.

One of our hopes in examining the simulated images was to help
constrain the inclination angle. The observations shown in
Fig.~\ref{fig:147_model_images} are for an inclination of
$0^\circ$. At an inclination angle of $30^\circ$, the synchrotron
emission is spread out over a much larger area and has a much lower surface
brightness than in the observations for $i=0^\circ$, and shown
in~\citet{Williams:1997}. As from the spectral modelling, this leads
us to believe that the inclination of the system is quite low, between
$0^\circ$ and $30^\circ$. This is contrary to the conclusions
of~\citet{Pittard:2002b} where larger inclination angles were required
to recover the extended X-ray emission. However, an alternative
explanation for this is that the stellar X-ray emission is extended on
larger scales than previously thought~\citep[see][]{Skinner:2002}.

\section{Summary and Future Directions}
\label{sec:summary}
In this first paper we have modelled the thermal and synchrotron 
emission from early-type binary systems with a strong wind-wind 
collision. We have used appropriate simplifying assumptions to make
this initial investigation tractable. In particular, we have generated models
of wide systems without considering IC cooling, and have assumed that the
relativistic particle energy distribution is a power-law up to 
infinite energies and is spatially invariant throughout the 
shocked gas. We have also assumed that the magnetic field is highly
tangled so that the synchrotron emission is isotropic.

We have demonstrated the importance of considering extended emission
and absorption regions, as opposed to treating these in a point-like
manner as in previous work. We have also demonstrated the importance
of synchrotron self-absorption, the Razin effect, and the free-free opacity
of the shocked gas in these systems. With our rather simple 
model we have been able to model both
the spectrum and the spatial distribution of radio emission from
\object{WR\,147} remarkably well. However, our simple model fails to address the
apparent high frequency cut-off in the synchrotron spectrum, due to
the current lack of a high energy cutoff in the relativistic electron
energy distribution. We note that this cutoff, which suggests a
characteristic value of $\gamma \sim 1200$, can be naturally explained
by IC cooling.

Clearly if we are to realistically model the observed radio emission
from CWBs we will need to calculate the evolution of the relativistic
particle energy spectrum as the particles are advected
downstream. This scenario will be addressed in a follow-up paper
(Pittard \etal, in preparation) where we plan to introduce a
finite maximum energy for the relativistic particles and allow this to
change along the shock front, as expected for variable IC cooling by
photons from the OB star. The inclusion
of IC cooling will enable the calculation of the evolution of the
energy spectrum of the relativistic particles downstream of the shock
acceleration region. Similarly the effect of Coulomb cooling on the
lower energy relativistic particles will be incorporated. The emission
spectrum of mono-energetic particles can then be convolved with the
energy spectrum of the particles to determine the total emission.

With these additional mechanisms it will be possible to investigate
the high frequency cut-off, such as that apparent in \object{WR\,147}, and to
more accurately model the low frequency cut-off when coulombic cooling
is important. Addressing these processes is vital if we are to model
smaller binary systems where the electron energy spectrum is highly
variable, largely due to the IC cooling from stellar UV photons. This
is certainly the case in the proto-typical CWB system \object{WR\,140}, where it
is clear the relative timescales for shock acceleration of electrons,
IC cooling, coulombic cooling and adiabatic expansion are varying 
dramatically throughout the 7.9-yr orbit. At the very least, these effects, 
along with radiative shock cooling, need to be considered in any 
realistic model of \object{WR\,140}.

\begin{acknowledgements}
We would like to thank John Dyson, Tom Harquist, Tony Moffat and Andy
Pollock for stimulating discussions related to this work, {and the
anonymous referee for raising several interesting issues arising in our
original manuscript.}  SMD would like to thank the University of Leeds,
UK for their hospitality during a number of visits, and JMP is
grateful for hospitality received at DRAO, Canada. JMP is supported by
a PDRA position from PPARC, and LK has been supported by the NRC Women
in Science and Engineering Program.  This research has made use of
NASA's Astrophysics Data System Abstract Service.
\end{acknowledgements}
\bibliography{theory_paper}

\appendix
\section{Radiative Transfer}
\noindent The equation of radiative transfer in a medium in which emitting and absorbing material 
is present, in the absence of scattering, can be written
\begin{equation}
\frac {dI_{\nu}}{ds} = \varepsilon_{\nu} - \alpha_{\nu} I_{\nu}
\end{equation}
where $\varepsilon_{\nu}$ and $\alpha_{\nu}$ are the emission and absorption coefficients at
frequency $\nu$, $s$ is the co-ordinate along a line of sight, and $I_{\nu}$ is the intensity
of the radiation field. This has a solution, at a given frequency ${\nu}$
\begin{equation}
I (s) = I (0)e^{-\tau (s)} + \int_0^s \varepsilon  (s^{\prime}) e^{-\tau(s^{\prime} \rightarrow s)}
ds^{\prime}
\label{eq:rad_transfer}
\end{equation}
where 
\begin{equation}
\tau (s) = \int_0^s \alpha (s^{\prime}) ds^{\prime}{~\rm and~~}
\tau (s^{\prime} \rightarrow s) = \int_{s^{\prime}}^s \alpha (s^{\prime \prime}) ds^{\prime \prime} .
\end{equation}
Consider a line of sight along which $\alpha$ and $\varepsilon$ are piecewise 
constant, that is 
\begin{equation}
\varepsilon = \varepsilon_i \mbox{    and     } \alpha = \alpha_i \mbox{    for    } 
(i-1)\Delta s \leq s < i \Delta s
\end{equation}
where $\Delta s$ is a finite increment of path length and $i = 1 \ldots n$.
Solving the integrals in the solution 
to the radiative transfer equation (Eqn.~\ref{eq:rad_transfer}) 
we can write the emergent intensity from an increment of path length
$(i-1)\Delta s \leq s < i \Delta s$ as follows

\begin{equation}
I_i = I_{i-1}e^{-\alpha_i \Delta s} + \int_{(i-1)\Delta s}^{i\Delta s}
\varepsilon_i e^{-\alpha_i (i\Delta s - s^{\prime})} ds^{\prime}
\end{equation}

\begin{equation}
I_i = I_{i-1}e^{-\alpha_i \Delta s} + \varepsilon_i 
e^{- \alpha_i i\Delta s} \int_{(i-1)\Delta s}^{i\Delta s}
\varepsilon_i e^{-\alpha_i s^{\prime}} ds^{\prime}
\end{equation}

\begin{equation}
I_i = I_{i-1} e^{-\alpha_i \Delta s} + \frac{\varepsilon_i}{\alpha_i} \left (
1 - e^{-\alpha_i \Delta s} \right ).
\label{eq:radt}
\end{equation}
Starting with $I_1$, and applying this scheme to each increment of path length in turn, 
we can derive the emergent intensity $I_n$ along the line of sight. 
In general, the increment of path length $\Delta s$ need not be constant,
as long as the emission and absorption coefficients are constant within
a given $\Delta s$.

This scheme forms the basis of the code.  Images are derived by
calculating the emergent intensity on a number of lines of sight which
pass through a grid of emission and absorption coefficients defined in
axis-symmetric cylindrical polar coordinates.  The geometry of the
problem is shown in Fig.~A.1. It is assumed, without any loss of
generality, that the line of sight which passes through the origin of
the coordinates on the model grid also passes through the origin of
the coordinates on the plane of the sky, and that the projection of
the line of sight onto planes of constant $z$ is parallel to the line
$\phi =0$.
\begin{figure}
\label{fig:geometry}
\vspace{14cm}
\includegraphics{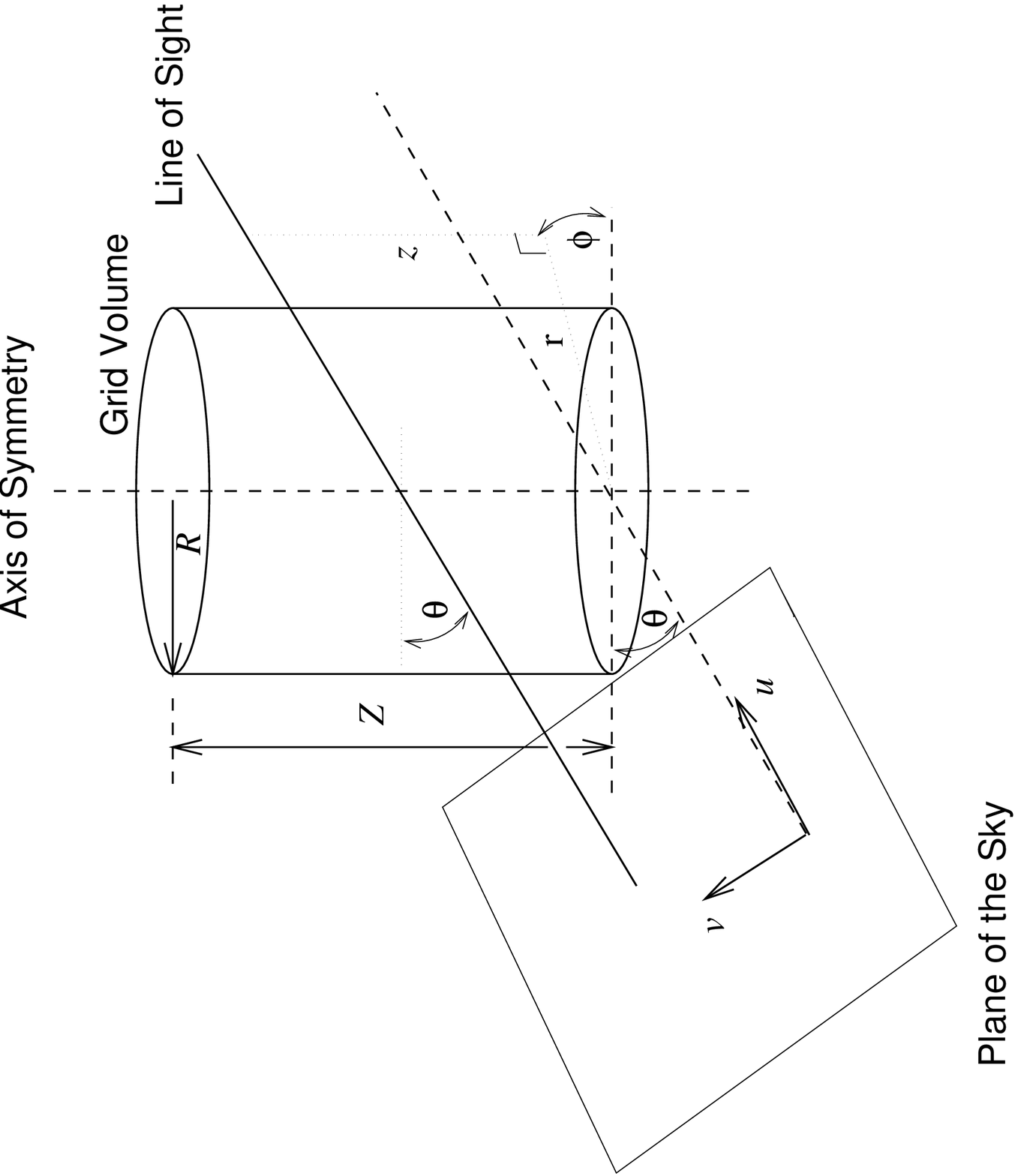}
\includegraphics{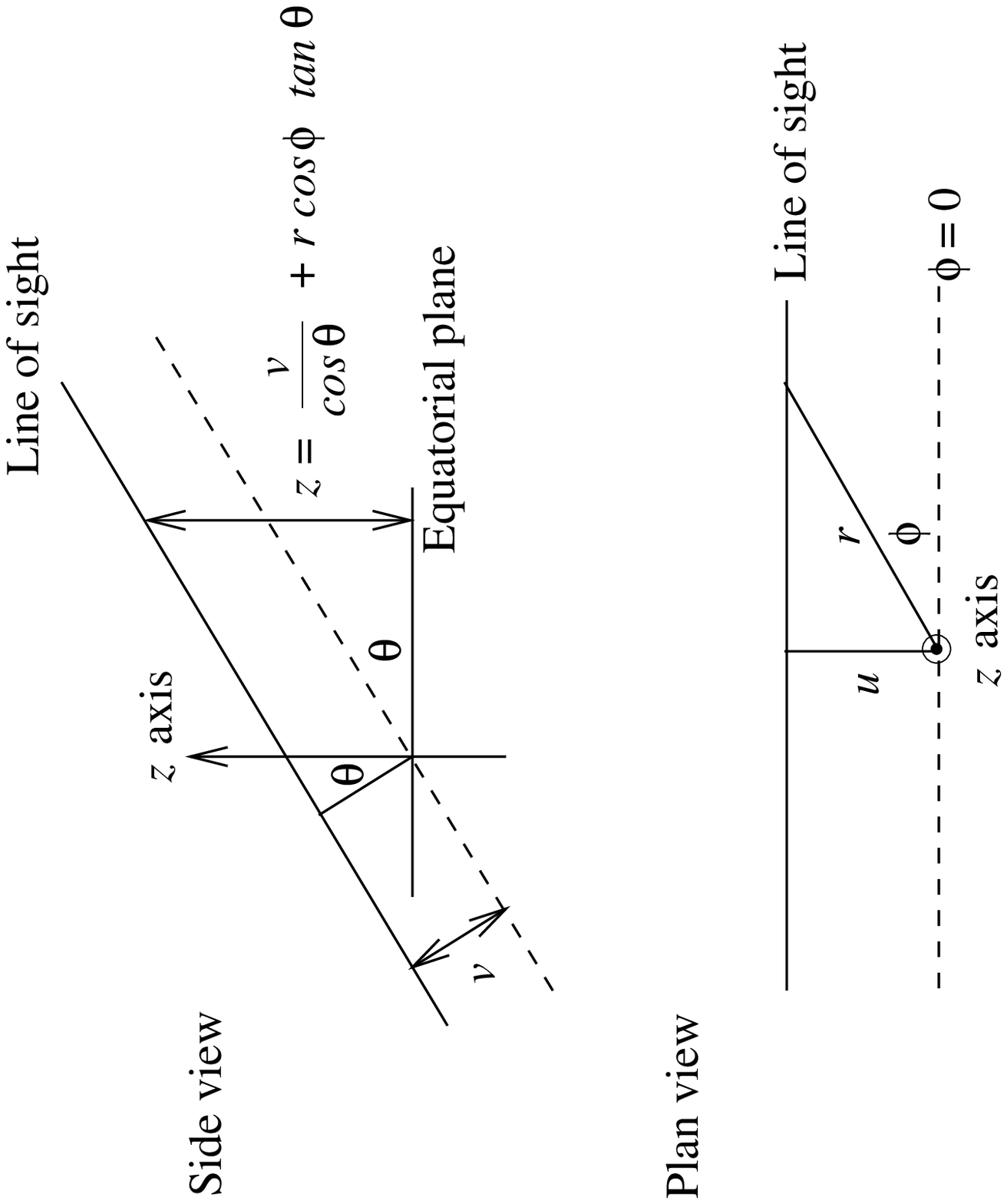}
\caption{Geometry for solving a line of sight through a grid in cylindrical polar
coordinates}
\end{figure}

The cell $j,k$ on the model grid is bounded by 
\begin{equation}
(j-1)\Delta r\leq r < j\Delta r \mbox{    and    } (k-1)\Delta z\leq z < k\Delta z 
\label{eq:arz}
\end{equation}
where $\Delta r$ and $\Delta z$ are, respectively, the width and height of a cell in the
$r,z$ plane. The grid consists of $N_r$ cells in the $r$ direction by $N_z$ cells
in the $z$ direction.
The domain covered by the grid is therefore 
\begin{equation}
r < R ; 0 \leq z < Z
\end{equation}
where $R = N_r \Delta r$ and $Z = N_z \Delta z$.
The $z$ axis of this grid is inclined at an angle $\theta$ to the plane
of the sky, on which we define a co-ordinate system $u, v$.
A line of sight through the grid is therefore defined by $u, v$ and $\theta$.
All lines of sight are normal to the $u, v$ plane.

It is convenient to define a point along a line of sight in terms of its 
azimuthal angle $\phi$ in the frame of the grid of emission and absorption coefficients. 
The co-ordinates on this grid of
the point $\phi$ along the line of sight $u, v, \theta$ are given
by
\begin{eqnarray}
r & = & \frac{u}{\sin \phi} \label{eq:r}\\
z & = &\frac{v}{\cos \theta} + u \frac{\tan \theta}{\tan \phi}.\label{eq:z}
\end{eqnarray}
The point, $\phi_0$, at which the line of sight enters the grid is given by 
\begin{equation}
\phi_0 = \mbox{min} \left\{ \sin^{-1} \frac{u}{R} , \tan^{-1} \left 
( -\frac{u}{v} \sin \theta \right) \right \}.
\end{equation}
The value of $\phi_0$ can be used to determine the indices ($j,k$) of the first
grid cell which encounters the line of sight, using Eqn.~\ref{eq:arz},~\ref{eq:r}, 
and~\ref{eq:z}. The indices of the next cell are determined from the minimum of $\phi_r$ and 
$\phi_z$, which are the values of $\phi$ corresponding to the 
boundaries with the next cells in the $r$ and $z$ directions respectively. 
The path length through the current cell is then given by
\begin{equation}
\Delta s = u \sqrt{1 + \tan^2 \theta} \left ( \frac{1}{\tan \phi_1} - \frac{1}{\tan 
\phi_{_{0}}} 
\right )
\end{equation}
where $\phi_1$ is the point at which the line of sight crosses the next cell boundary.

For a given line of sight, we now have the grid indices of the first cell
which encounters the line, the indices of the next cell, and the path
length in the current cell.  Having obtained the emission and
absorption coefficients from the appropriate grid cell, we can update
the intensity on the given line of sight using Eqn.~\ref{eq:radt}. This
process is repeated for a line of sight until it leaves the grid (i.e.
if $i > N_r$ or $j > N_z$).  

\label{lastpage}

\end{document}